\documentclass[aps,prl,twocolumn,tightenlines,superscriptaddress,showpacs]{revtex4-1}
\usepackage{epsfig,rotating,amsmath,lineno}
\usepackage{multirow}

\begin{document}

\title{The nuclear shell model towards the drip lines}

\author{B. Alex Brown}
\affiliation{Department of Physics and Astronomy, and
Facility for Rare Isotope Beams,
Michigan State University, East Lansing,
Michigan 48824-1321, USA}

\begin{abstract}

Applications of configuration mixing methods for nuclei near the
proton and neutron drip lines are discussed.
A short review of magic numbers is presented.
Prospects for
advances in the regions of four new "outposts" are highlighted:
$^{28}$O, $^{42}$Si, $^{60}$Ca and $^{78}$Ni. Topics include: shell gaps,
single-particle properties, islands-of-inversion,
collectivity, neutron decay, neutron halos, two-proton decay,  effective charge,
and quenching in knockout reactions.
\end{abstract}

\maketitle

\section{Introduction}
The starting point for the nuclear shell model is the establishment
of model spaces that allow for tractable configuration interaction (CI) calculations
from which we are able to understand and predict the properties of low-lying states
\cite{usd}, \cite{br01}, \cite{ca05}, \cite{st19}, \cite{ot20}.
This choice is based on the observation that a few even-even nuclei can be
interpreted in terms of having magic numbers for $  Z  $ or $  N  $ and doubly-magic
numbers for a given $  (Z,N)  $.
These magic numbers can be inferred from
experimental excitation energies of 2$^{ + }$ states shown for the low
end of the nuclear chart in Fig. (1).
Magic numbers are those values of $  Z  $ or $  N  $ for
nuclei that have a relatively high 2$^{ + }$ energy within a series of isotopes or isotones.

\begin{figure}
\includegraphics[scale=0.45]{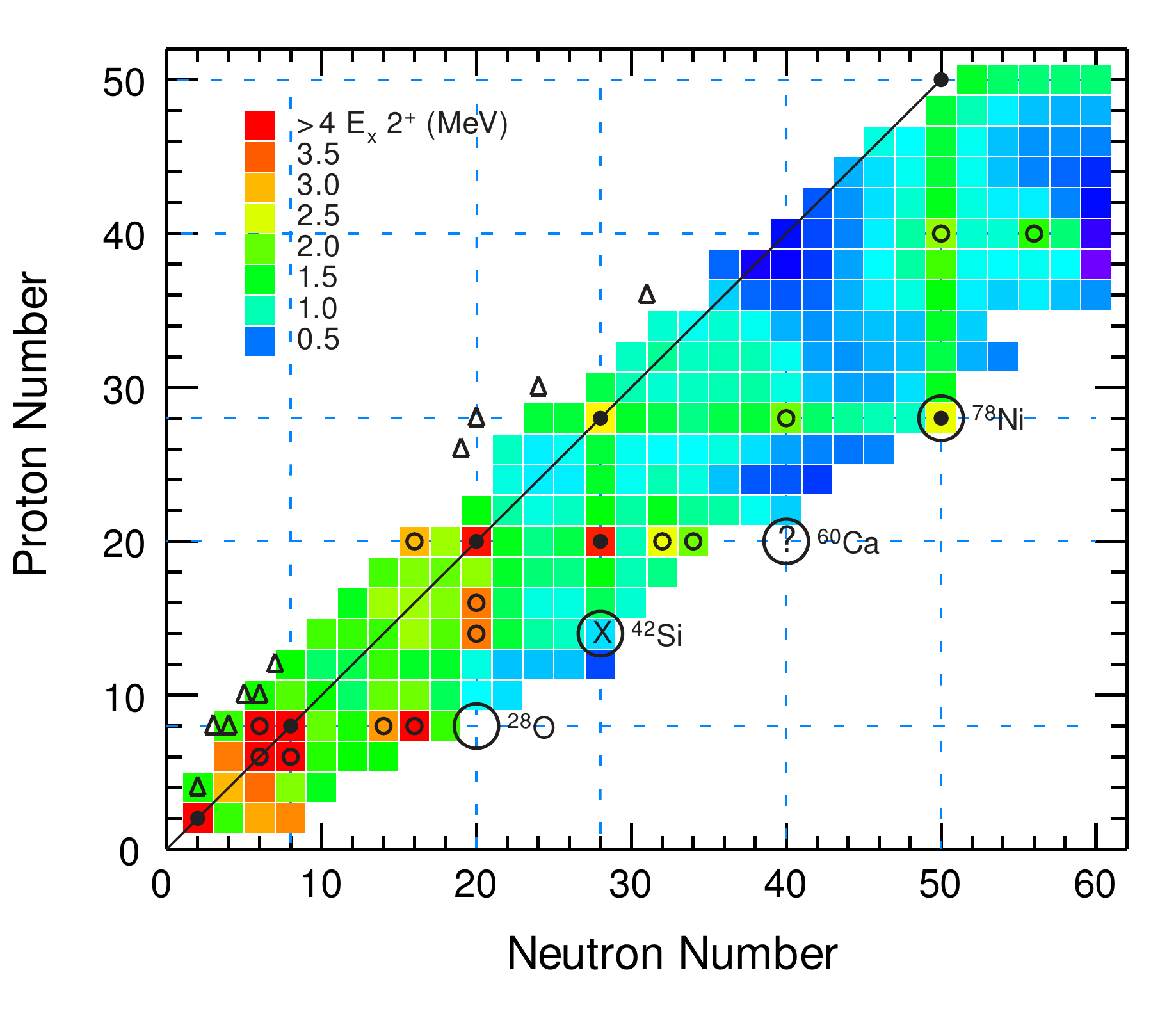}
\caption{Lower mass region of the nuclear chart.
The colors indicate the energy of the first 2$^{ + }$ state.
In addition to the data from \cite{pr16}, we add
recent data for $^{40}$Mn \cite{cr19}, $^{62}$Ti \cite{cortes20},
$^{66}$Cr \cite{sa15} and $^{70,72}$Fe \cite{sa15}.
The filled black circles show the doubly-magic nuclei
associated with the most robust pairs of magic numbers 8, 20, 28  and 50.
The open circles show the doubly-magic nuclei associated with
less robust magic numbers 14, 16, 32, 34 and 40.
The large open circles indicate the nuclei near the neutron
drip lines that are the focus of this paper.
The triangles are those nuclei observed to decay by two protons in
the ground state.
}
\label{ (1) }
\end{figure}

Another measure of magic numbers is given by the
double difference in the BE defined by
$$
D(q) = (-1)^{q} [2 {\rm BE}(q) - {\rm BE}(q+1) + {\rm BE}(q-1)],       \eqno({1})
$$
for isotopes ($  q=N  $ with $  Z  $ held fixed) or isotones ($  q=Z  $ wth $  N  $ held fixed)
can also be used to measure shell
gaps \cite{br13}.
An example for the neutron-rich calcium isotopes is
shown in Fig. (2).
(The dashed line extrapolation to $  N=40  $ will be discussed below.)
The value of $  D(N)  $ at these magic numbers gives the effective shell gap.
In between the magic numbers, $  D(N)  $ gives the pairing energy \cite{br13}.
The excitation energies of the
2$^{ + }$ states at $  N=28  $, 32 and 34, also shown in the figure, are
close to the $  D(N)  $ values at these magic numbers.
The neutron gaps at $  N=32  $ and 34 are weaker than the
gap at $  N=28  $, but they are strong enough to allow the
configurations to be dominated by the orbitals shown in Fig. (2).

In the simplest model, the magic number is associated with a ground state
that has a closed-shell configurations for the given value of $  Z  $ or $  N  $.
The following is from footnote (9) of Ref. \cite{au06}.
It was Eugene Paul Wigner who coined the term "magic number".
Steven A. Moszkowski, who was a student of Maria Goeppert-Mayer, in a talk
presented at the APS meeting in Indianapolis, May 4, 1996 said:
"Wigner believed in the liquid drop model, but he recognized, from the work of Maria
Mayer, the very strong evidence for the closed shells. It seemed a little like
magic to him, and that is how the words ‘Magic Numbers’ were coined.”
The discovery of
"magic numbers”, lead M. Goeppert-Mayer, and independently J. Hans
D. Jensen in Europe, 1 year later in 1949, to the construction of the shell model
with strong spin-orbit coupling, and to the Nobel prize they shared with Wigner
in 1963.

The nuclei marked with closed circles in Fig. (1) are commonly
used to define the boundaries of CI model spaces. Those indicated by open circles
are usually contained within a larger CI model spaces.
Historically, the size of the assumed model space has depended on the computational capabilities.
At the very beginning in the 1960's they were the $  0p  $ model space bounded by $^{4}$He and 
$^{16}$O, and
the $  0f_{7/2}  $ model space bounded by $^{40}$Ca and $^{56}$Ni.

For heavy nuclei, doubly-magic nuclei are associated with the
shell gaps at 28, 50, 82 and 126. These are created by the spin-orbit
splitting of the high $\ell$ orbitals which lowers the
the $  j=\ell +1/2  $ single-particle energies for $\ell$=3 (28), $\ell$=4 (50),
$\ell$=5 (82) and $\ell$=6 (126). I will refer to these as $  jj  $ magic numbers.
The nuclei with $  jj  $ magic numbers for both protons and neutrons
will be called double-$  jj  $ closed-shell nuclei.
These are shown by the red circles in Fig. (3):
$^{208}$Pb, $^{132}$Sn, $^{100}$Sn, $^{78}$Ni and $^{56}$Ni.
The open red circle for $^{100}$Sn indicates that it is expected
to be double-$  jj  $ magic \cite{mo18}, but it has not yet been confirmed experimentally.
The continuation of the double-$  jj  $ sequence with $\ell$=2 (14) and $\ell$=1 (6)
are shown by the  open blue circles for
$^{42}$Si, $^{28}$Si, $^{18}$C and $^{12}$C on the lower left-hand side of Fig. (3).
As I will discuss below, the calculations for these nuclei show
rotational bands with positive quadrupole moments indicative of an oblate intrinsic shape.

In light nuclei, magic numbers are associated with the filling of a
major harmonic-oscillator shell with $  N_{o}=(2n+\ell )  $ where both members of the
spin-orbit pair $  j=\ell  \pm 1/2  $ are filled. I will refer to these
as $  LS  $ magic numbers.
The $  LS  $ magic numbers for isotopes and isotones are shown by the
thin brown lines in Fig. (4).
There are only three known double-$  LS  $ magic nuclei,
$^{4}$He, $^{16}$O and $^{40}$Ca shown by the filled red circles in Fig. (4).
The next one in the sequence would be $^{80}$Zr, but in this case
the $  Z=N=40  $ gap is too small due to the lowering of
the $  0g_{9/2}  $ single-particle energy due to the spin-orbit splitting.
As will be discussed below,
$^{60}$Ca (the red open circle with a question mark) could be a "fourth" double-$  LS  $
magic nucleus.
There are regions where the $  LS  $
magic numbers for isotopes or isotones dissappear as shown by the
blue lines in Fig. (4). These will be referred to as
"islands-of-inversion" \cite{wa90}.

The nuclei with green circles in Fig. (4) also have doubly-magic properties.
The pattern is that when one kind of nucleon (proton or neutron)
has an $  LS  $ magic number, then
the other kind of nucleon has a magic number for the filling or each
$  j  $ orbital.
These are 6 ($  0p_{3/2}  $), 8 ($  0p_{1/2}  $), 14 ($  0d_{5/2}  $), 16 ($  1s_{1/2}  $),
20 ($  0d_{3/2}  $),
28 ($  0f_{7/2}  $), 32 ($  1p_{3/2}  $), 34 ($  1p_{1/2}  $), 40 ($  0f_{5/2}  $),
50 ($  0g_{9/2}  $) and 56 ($  1d_{5/2}  $).

The only addition to the $  jj  $ and $  LS  $ closed-shell systematics
discussed above is for $^{88}$Sr shown in Fig. (4) where there
is an energy gap between the proton $  1p_{1/2}  $ and $  1p_{3/2}  $,$  0f_{5/2}  $
states. In early calculations $^{88}$Sr was used as the closed shell
for the $  1p_{1/2},0g_{9/2}  $ model space \cite{se76}, but more recently
the four-orbit model space of $  0f_{5/2}  $,$  1p_{3/2}  $,$  1p_{1/2}  $,$  0g_{9/2}  $
has been used for the $  N=50  $ isotones \cite{ji89}, \cite{li04}.

For a given shell gap, the $  LS  $ magic
numbers are more robust than those for $  jj  $. The reason is that
deformation for $  jj  $ magic numbers starts with a one-particle
one-hole ($  1p-1h  $) excitation of a nucleon in the $  j=\ell +1/2  $ orbital to the
other members of the  same oscillator shell, $  N_{o}=(2n+\ell )  $.
Since $  1p-1h  $ excitations across $  LS  $ closed shell gaps change parity,
ground-state deformation for $  LS  $ magic numbers
must come from $  np-nh  $ ($  n \geq 2  $) excitations
across the $  LS  $ closed shells as in the region
of $^{32}$Mg \cite{wa90}.

\begin{figure}
\includegraphics[scale=0.6]{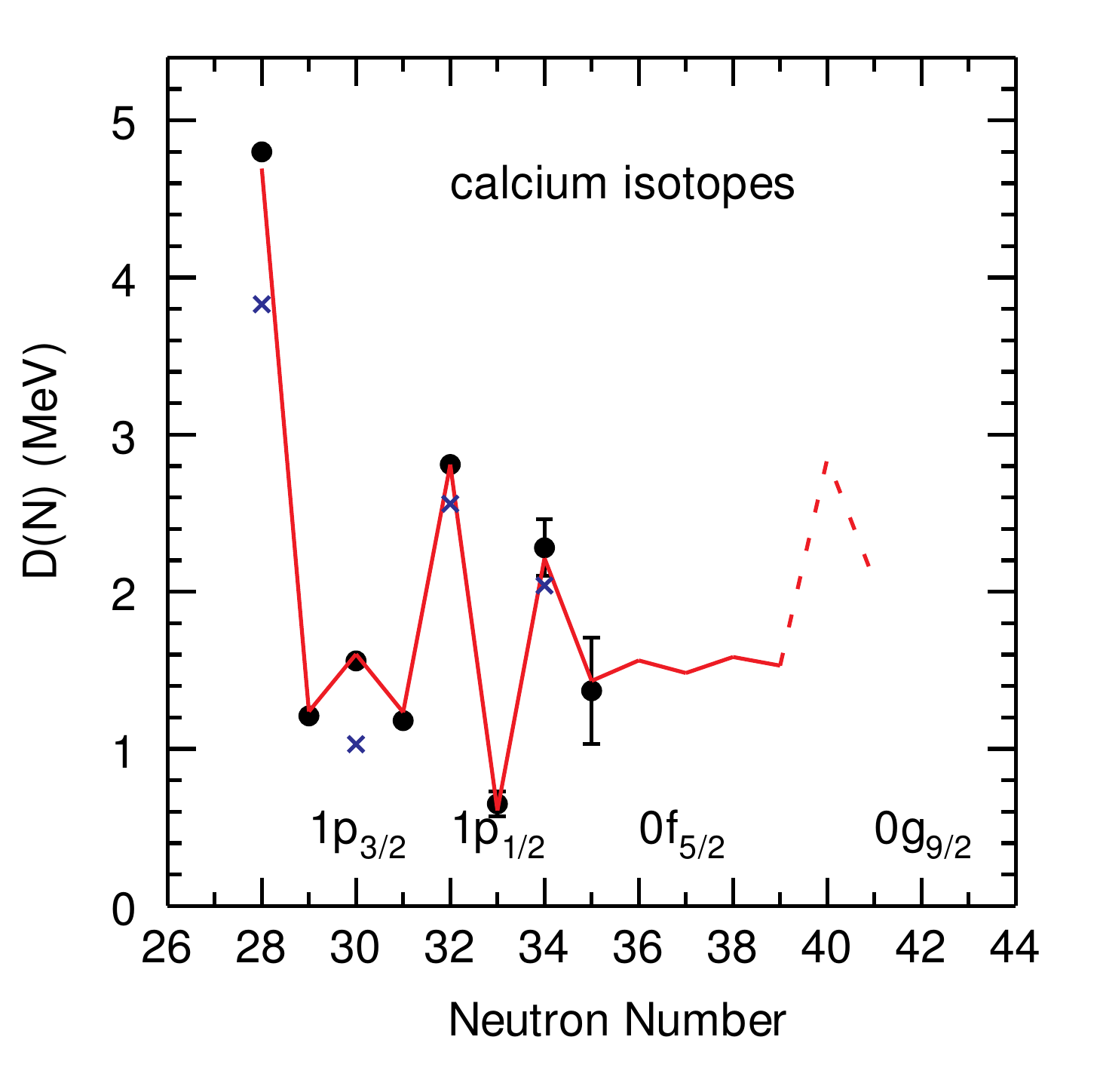}
\caption{$  D(N)  $ as given by Eq. (1). The black dots with error bars
are the experimental data. The blue crosses are the excitation energies
of the 2$^{ + }_{1}$ states. The orbitals that
are being filled are shown.
The red line is the results from the UFP-CA
Hamiltonian \cite{ma21}. The dashed line is the extrapolation based on the
UNEDF0 BE for $^{60,61,62}$Ca \cite{unedf1}.
}
\label{ (2) }
\end{figure}

\begin{figure}
\includegraphics[scale=0.5]{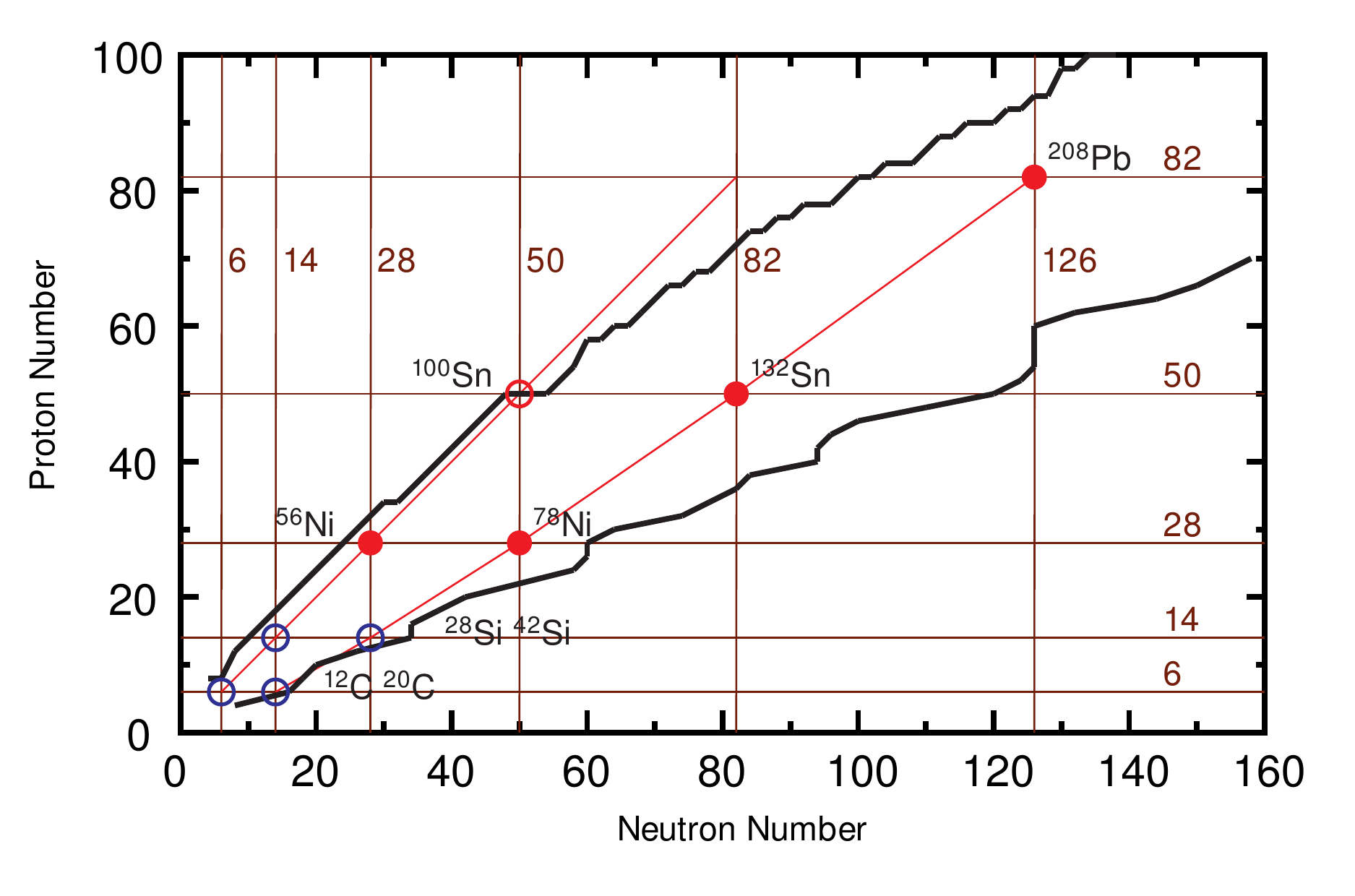}
\caption{The nuclear chart
showing the $  jj  $ magic numbers.
The black lines show where the two-proton (upper)
and two-neutron (lower)
separation energies obtained with the UNEDF1 \cite{unedf1}
cross one MeV.
The filled red circles show the locations of double-$  jj  $ magic nuclei
established from experiment.
The open red circle for $^{100}$Sn indicates a probably double-$  jj  $
magic
nucleus that has not been confirmed by experiment.
The blue circles in the bottom left-hand side are nuclei
in the double-$  jj  $ magic number sequence that
are oblate deformed.
}
\label{ (3) }
\end{figure}

\begin{figure}
\includegraphics[scale=0.6]{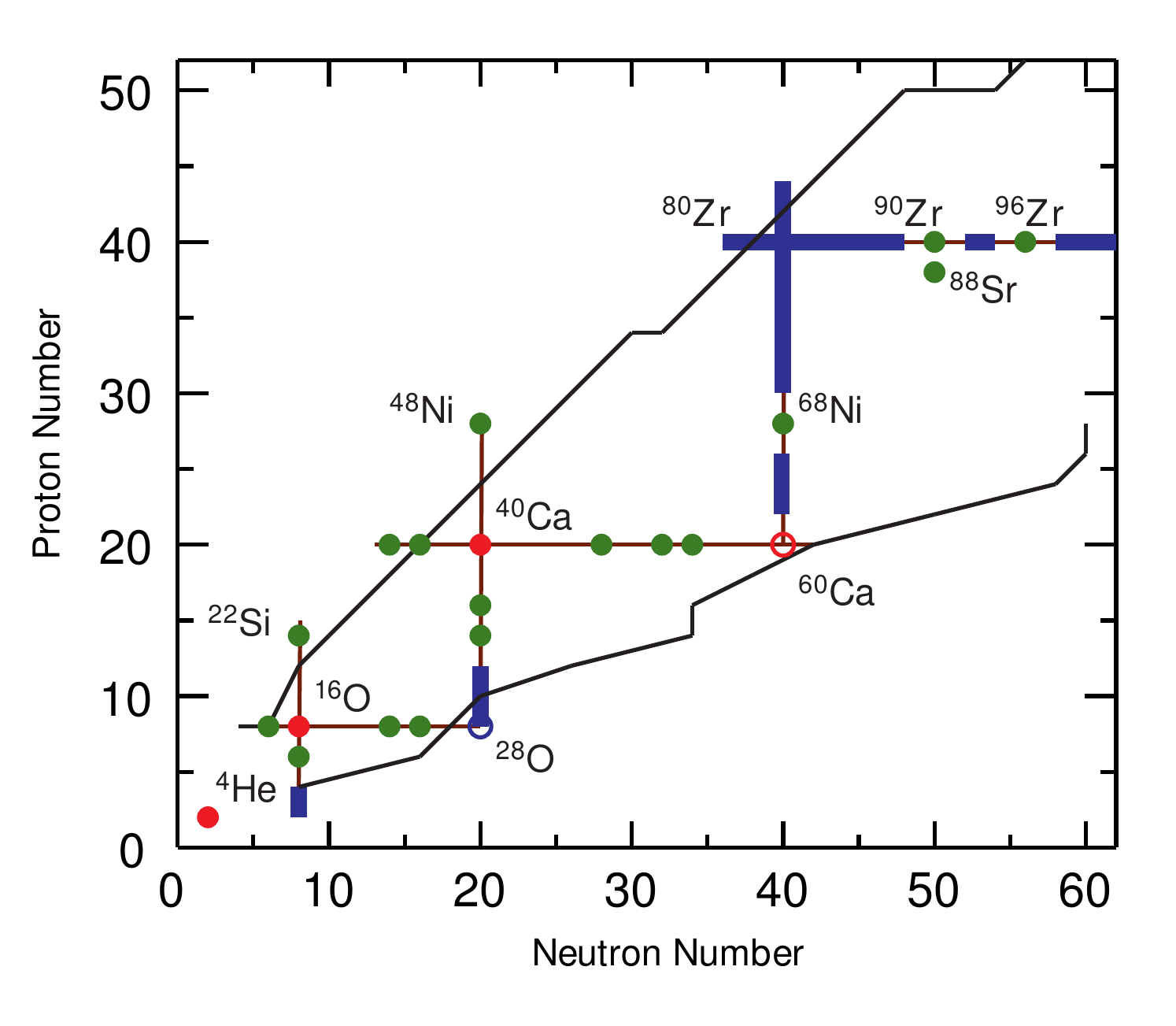}
\caption{Lower mass region of the nuclear chart
showing the $  LS  $ magic numbers, 2, 8, 20 and 40.
The black lines show where the two-proton (upper)
and two-neutron (lower)
separation energies obtained with the UNEDF1 \cite{unedf1}
functional cross one MeV.
The filled red circles show the double-$  LS  $ magic nuclei
$^{4}$He, $^{16}$O and $^{40}$Ca.
The open red circle for $^{60}$Ca indicates a possible doubly-magic
nucleus that has not been confirmed by experiment.
The green circles are doubly-magic nuclei associated
with the $  j  $-orbital fillings.
The blue lines indicate isotopes or isotones where the $  LS  $
magic number is observed to be broken.
}
\label{ (4) }
\end{figure}

I will discuss results obtained with Hamiltonians
based on data-driven
improvements to the two-body matrix elements provided by ab-initio methods.
The ab-initio methods are based on NN and NNN interactions
obtained by model-dependent fits to nucleon-nucleon phase shifts
and properties of nuclei with $  A=2-4  $. For a given model
space, these are renormalized for short-range correlations
and for the truncations into the chosen model space
to provide a set of two-body matrix elements (TBME)
for nuclei near a chosen doubly-closed shell.
From this starting point, one attempts
to make minimal changes to the Hamiltonian to improve
the agreement with energy data for a selected set
of nuclei and states within the model space.  A convenient way to
do this is done by using the singular value decomposition (SVD) method \cite{usdc}.
In many cases one adjusts specific TBME or combinations
of TBME. The most important
are the monopole, pairing and quadrupole components.
An important part of
the universal Hamiltonian is in the evolution of the effective
single-particle energies (ESPE) as one changes the number of protons and neutrons.
Starting with a closed shell with a given set of single-particle
energies, the ESPE as a function of $  Z  $ and $  N  $
are determined by the monopole average parts of the TBME \cite{ot20}.

These methods provide "universal" Hamiltonians
in the sense that a single set of single-particle energies and two-body matrix elements
are applied to all nuclei in the model space, perhaps
allowing for some smooth mass dependence. This has turned out to be
a practical and useful approximation. As the ab-initio starting points are improved,
these "universal" Hamiltonians will be replaced by Hamiltonians for
a more restricted set of nuclei, or even for individual nuclei
as has been done in the VS-IMSRG method \cite{st19}, \cite{st21}.

The empirical modifications to the effective Hamiltonian
account for deficiencies in the more ab-initio methods.
Most ab-initio calculations are carried out in a harmonic-oscillator basis
due to its convenient analytical properties.
Near the neutron drip lines the radial wavefunctions
become more extended, the single-particle energy spectrum becomes
more compressed, and the continuum becomes explicitly more important.
To take this into account,
the ab-initio methods require a very large harmonic-oscillator
basis.

Due to the continuum, nuclei near the neutron drip line
present a substantial theoretical challenge \cite{br01}, \cite{fo13}.
Methods have been developed that take the continuum into account
explicitly.
The density matrix renormalization
group (DMRG) method \cite{ro06}, \cite{ro09} makes use
of a potential together with a simplified interaction
based on halo effective field theory \cite{be02}, \cite{be03}.
In the Gamow shell model (GSM) \cite{id02}, \cite{mi02}, \cite{mi09}, the many-body basis is
constructed from a single-particle Berggren ensemble \cite{be93}, \cite{li93}.
The DMRG and GSM  methods rely on use of simplified two-body interactions
with adjustable parameters. There is also the shell-model embedded in
the continuum formalism that can make use of the universal interactions \cite{vo05}.
Recent progress in the GSM method is presented in another contribution to
this series \cite{li21}.

Ground-state nuclear halos are a unique feature of nuclei near the neutron drip line
\cite{ta96}.
This is due to the loose-binding of low-$\ell$ orbitals with extended radial
wavefunctions.
The most famous case is that for $^{11}$Li which was observed to
have a rapid rise in the nuclear matter radius compared to the trends up to $^{9}$Li \cite{ta85}.
The wavefunction of $^{11}$Li is dominated by a pair of neutrons in the $  1s_{1/2}  $ orbital.
As discussed below, halos in the region of $^{30}$Ne and $^{42}$Si are dominated by the $  1p_{3/2} 
 $
orbital. Proton halos are not so extreme due to the Coulomb barrier.
The excited 1/2$^{ + }$ ($  1s_{1/2}  $) state of $^{17}$F is a good example of
an excited-state halo as determined indirectly
from its large Thomas-Ehrman energy shift of 0.87 MeV $^{17}$O
to 0.49 MeV in $^{17}$F.

States above the (proton/neutron) separation energy have
(proton/neutron) decay widths.
In the conventional CI approach, one calculates states whose
energy is taken to be the centroid energy of the decaying state.
The decay width is calculated using the approximation
$  \Gamma  = {\rm C^{2}S} \, \Gamma _{sp} (Q)  $ where C$^{2}$S is the spectroscopic factor
and $  \Gamma _{sp}  $ is the single-particle neutron decay width calculated with
a $  Q  $ value taken from the shell-model centroid or the experimental
centroid if known. The explicit addition of the continuum
shifts down the energy relative to its CI energy \cite{vo05}.
Also, the continuum (finite-well potential) is responsible for
the Thomas-Ehrman shift
for states in proton-rich nuclei
compared to those in the neutron-rich mirror nuclei \cite{usdc}.

In this review I  will concentrate on four regions of
neutron-rich "outposts"
whose understanding are most important for future developments.
These are shown in Fig. (1): $^{28}$O, $^{42}$Si, $^{60}$Ca and $^{78}$Ni.
$^{42}$Si is labeled by "x" since it does not have a magic number
for protons or neutrons. $^{78}$Ni is labeled by a filled circle
since it is now known to be doubly magic \cite{ni78}.
$^{60}$Ca is known to be inside the neutron drip line \cite{ca60},
but its mass and excited states have not yet been measured.

Nuclei that are observed
to decay by two protons are shown by the triangles in Fig. (1).
The two-proton ground-state decays for $^{45}$Fe, $^{48}$Ni, $^{54}$Zn and $^{67}$Kr
have half-lives on the order of ms and compete with the $\beta$ decay
of those nuclei. An experimental and theoretical summary of the results
for those nuclei together with that of $^{19}$Mg has been given in \cite{br19}.
There is qualitative agreement between experiment and theory.
In order to become more quantitative, the experimental errors in the
partial half-lives need to be improved.
Theoretical models need to be improved to incorporate
three-body decay dynamics (presently based on single-orbit configurations)
with the many-body CI calculations for the two-nucleon decay amplitudes.
The correlations for
two-nucleon transfer amplitudes via (t,p) or ($^{3}$He,n) are largely determined
by the $  (S,T)=(0,1)  $ structure of the triton or $^{3}$He, whereas two-nucleon
decay is determined by the decay through the Coulomb and angular-momentum
barriers that are dominated by the low-$\ell$ components.
For the lightest nuclei, multi-proton emissions (shown in Fig. 1 of
\cite{ji21}) are observed as
broad resonances.

Knockout reactions are used to produce nuclei further from stability.
The cross sections for these reactions can be compared to
theoretical models in terms of the cross-section ratio $  R_{s}=\sigma _{exp}/\sigma _{th}  $.
See \cite{to21} for a recent summary. It is observed for
nuclei far from stability where $  \Delta S=\,\mid S_{1p}-S_{1n}\mid   $ is large
($  S_{1}  $ is the one nucleon separation energy) that
$  R_{s}  $ is near unity when the knocked out nucleon is loosely bound
but drops to about 0.3 for deeply bound nucleons.
This has been attributed to the
short- and long-ranged correlations that depletes
the occupation of deeply-bound states \cite{pa20}.
The short-ranged correlations are connected to the high-momemtum
tail observed in observed in high-energy electron scattering
experiments \cite{he17}.
The long-ranged correlations come fron particle-core coupling
and pairing correlations beyond that included within the valence space.
Another reason may be the approximations made in the sudden approximation
for the dynamics used for the reaction \cite{to21}.
In the analysis of \cite{pa20}, the $  R_{s}  $ factor for loosely-bound
nucleons that comes mainly from the long-ranged correlations
is expected to be 0.6-0.7 rather than unity.
The analysis of $  (p,2p)  $ experiments \cite{pa16}
find $  R_{s}  $ values that depend less on the proton
separation energy going from 0.6 to 0.7.

The $  \sigma _{th}  $ depends on the CI calculations for the spectroscopic
factors. An approximation that is made in CI calculations
is that only the change in configurations for the
knocked out nucleon contributes to the spectroscopic factor.
The radial wavefunctions for all other nucleons
in the parent and daughter nuclei
are assumed to be the same. But consider, as an example,
the knockout of a deeply bound proton from $^{30}$Ne to $^{29}$F.
The size of the neutrons orbtials in $^{30}$Ne and $^{29}$F
are changing due to the proximity to the continuum,
and the overlap of of the spectator neutrons
in the nuclei with $  (A)  $ and $  (A-1)  $.
will be reduced from unity.
This effect should be contained
in ab-inito and continuum models \cite{je11}, \cite{wy21}, but
an understanding within these models requires an
explicit separation of the one-nucleon removal overlaps
in terms of the removed nucleon within the basis states
for $  (A,Z)  $ and the radial overlaps
between the nuclei with $  (A)  $ and $  (A-1)  $.

\section{The region of $^{28}$O}

The oxygen isotopes provided the first complete testing ground for theory and experiment
from the proton drip line to the neutron drip line \cite{ox17}. The prediction by the
USD Hamiltonian \cite{wi84}, \cite{usd}
in the 1980's that $^{24}$O was a doubly-magic nucleus was later
confirmed experimentally in 2009 \cite{ka09}, \cite{ho09}, \cite{ja09}.

For the one-neutron decay of $^{25}$O, the USDC Hamiltonian in the $  sd  $ shell \cite{usdc}
gives $  Q=1.15(15)  $ MeV, to be compared to the experimental value of $  Q=0.749(10)  $ MeV 
\cite{o25}.
The explicit addition of the continuum will lower the calculated energy \cite{vo05}.
The calculated value of
the spectroscopic factor is $  (25/24)^{2} C^{2}S(0d_{3/2})=1.01(1)  $ (the error comes
from the comparison of the four $  sd  $-shell Hamiltonians developed in \cite{usdc}).
For the calculated decay width one obtains $  \Gamma  = {\rm C^{2}S} \, \Gamma _{sp} (Q) = 75(1)  $ 
keV.
$  \Gamma _{sp} = 74(1)  $ keV is obtained using the experimental $  Q  $ value and a Woods-Saxon
potential. The experimental neutron decay width is $  \Gamma  = 88(6)  $ keV \cite{o25}.
The theoretical error in the width is probably dominated by the parameters of the
Woods-Saxon potential.

The measured masses of the Na isotopes \cite{th75} found more binding
near $  N=20  $ than could be accounted for by the pure $  \Delta =0  $ configurations.
(I use the notation $  \Delta =n  $ where $  n  $ is the number of neutrons excited
from $  sd  $ to $  pf  $.)
Hartree-Fock calculations \cite{ca75} showed that these mass anomalies
were associated with a large prolate deformation, where the
2$\Omega^{ \pi }$ [N,n$_{z}$,$\Lambda$]=1$^{-}$ [3,3,0] and 3$^{-}$ [3,2,1]  Nilsson
orbitals
from the $  fp  $ shell
cross the  1$^{ + }$ [2,0,0]  and 3$^{ + }$ [2,0,2] orbitals from the $  sd  $ shell
near $\beta$ = $+$0.3.
The anomaly was confirmed by $  \Delta =0  $, CI calculations \cite{wi80}, \cite{wi83}
where in \cite{wi80} it was called the "collapse of the conventional shell-model."
CI calculations that included $  \Delta =2  $ components \cite{po87}, \cite{wa90} showed that
nuclei in this region have ground state wavefunctions dominated by the $\Delta$=2 component.
This is due to
a weakened shell gap at $  N=20  $ below $  14  $, pairing correlations in the
$\Delta$=2 configurations,
and proton-neutron quadrupole correlations that give rise
to the Nilsson orbital inversion.
In \cite{wa90} the region of nuclei below $^{34}$Si involved in this inversion
was called the "island-of-inversion".

The Hamiltonian used in \cite{wa90}
was appropriate for pure $  \Delta =n  $ configurations.
This Hamiltonian was
modified to account for more recent data related to the energies of $  \Delta =1  $
and $  \Delta =2  $ configurations \cite{fsu}. As examples of the type of
predictions, I show results obtained with the FSU Hamiltonian
for $^{34}$Si in Fig. (5), $^{32}$Mg in Fig. (6),
and $^{29}$F in Fig. (7). All of these calculaitons were carried out
with NuShellX \cite{nushellx} code and allowed only for neutron
excitations from $  1s-0d  $ to $  1p-0f  $. Calculations in a full $  n\hbar \omega   $
basis with $  n>0  $
also require the addition of proton excitations from $  0p  $ to $  1s-0d  $ and proton
exicitations from $  1s-0d  $ to $  1p-0f  $.
In full $  n\hbar \omega   $  basis, the 1$\hbar\omega$ spurious states can be removed
with the Gloeckner-Lawson method \cite{lawson}. Comparison to calculations in the full $  n\hbar 
\omega   $
basis with the Oxbash code \cite{oxbash} show that the energies
are lowered relative to the $\Delta$ basis by up to about 200 keV.
This shows the $\Delta$=1,2 proton and proton-neutron components
are small compared to the $\Delta$=1,2 neutron components for the
low-lying states in these neutron-rich nuclei.
For nuclei with $  N \approx Z  $, removal of the spurious
states in the $  n\hbar \omega   $ basis is important.

The barrier between the $  \Delta =0  $ (spherical) $  \Delta =2  $ (deformed)
configurations reduces the mixing between the lowest energy states
of each configuration.
When one combines the $  \Delta =0  $ and $  \Delta =2  $ configurations in CI calculations,
the state that is dominated by $  \Delta =0  $
is pushed down in energy by the mixing with many $  \Delta =2  $ configurations
mainly due to the increase in the pairing energy.
If one were to start with the FSU Hamiltonian and add off-diagonal
TBME of the type $  <sd\mid V\mid fp>  $,
the components dominated by $\Delta$=0 would be pushed down in energy
due to this increase in pairing. But this results in a double-counting
since the $  sd  $ part FSU interaction is already implicitly renormalized
for the $  fp  $ admixtures.
In addition, to achieve convergence in the mixed wavefunctions
one has to add $\Delta$=4 and higher. This results in
large matrix dimensions.

When one mixes the $\Delta$ components, one has to modify
parts of the Hamiltonian that are diagonal in $\Delta$.
This is sometimes done
approximately
by changing the pairing strength in the $  J=0  $ $  T=1  $ two-body
matrix elements, so that the ground-state binding energies
agree with experimental values.
Hamiltonians that have been designed for mixed configurations
are SDPF-U-MIX \cite{sdpfumix} and SDPF-M \cite{sdpfm}, \cite{sdpfm2}.
Details about the modifications to SDPF-U to obtain SDPF-U-MIX are
given in the Appendix of \cite{sdpfumix}.
In the remainder of this section I will discuss some examples obtained
with the FSU Hamiltonian with pure $\Delta$ configurations. This provides a starting point for
more complete calculations with mixed $\Delta$ and those explicitly
involving the continuum.

The $  \Delta =0  $ ($  sd  $-shell) part of the FSU spectrum for $^{34}$Si [the green lines
in Fig. (5)] has a simple interpretation. The ground state
is dominated by the $  (0d_{5/2})^{6}  $ proton configuration. The
the 5.24 MeV 2$^{ + }$ and the 6.47 MeV 3$^{ + }$ states
are dominated by the $  (0d_{5/2})^{5}(1s_{1/2})^{1}  $ proton configuration.
In the two-proton transfer experiment from $^{36}$S \cite{s36}
a 2$^{ + }$ state at 5.33 is observed that can be interpreted
as two protons removed from $  (0d_{5/2})^{6}(1s_{1/2})^{2}  $
to make  $  (0d_{5/2})^{5}(1s_{1/2})^{1}  $. The $  (0d_{5/2})^{4}(1s_{1/2})^{2}  $ 0$+$ state
is predicted at 8.76 MeV. For the FSU Hamiltonian all of these predictions are
based on the USDB effective Hamiltonian \cite{usdb}. The
ESPE for the $  0d_{5/2}  $ and $  1s_{1/2}  $ proton states near $^{34}$Si
are determined from the binding energies of $^{33}$Al,
$^{34}$Si and $^{35}$P.
Above 2.5 MeV the level density
is dominated by the neutron $  \Delta =1  $ and $  \Delta =2  $ configurations.
The $  \Delta =1  $ states can be interpreted in terms of the
low-lying 3/2$^{ + }$ and 1/2$^{ + }$ $  1h  $ states of $^{33}$Si coupled
to the low-lying 7/2$^{-}$ and 3/2$^{-}$ $  1p  $ states of $^{35}$Si.
The state with maximum $  J^{\pi }  $ of 5$^{-}$ predicted at 5.12 MeV can be compared to the
proposed experimental 5$^{-}$ state at 4.97 MeV \cite{li19}.
The theoretical spectra from the mixed SDPF-U-Mix shown in
\cite{li19} is similar to the FSU unmixed spectrum in Fig. (5).

The FSU results for $^{32}$Mg are shown in Fig. (6).
Compared to $^{34}$Si there is an inversion of the low-lying $\Delta$=0
and $\Delta$=2 configurations.
For pure $\Delta$ configurations the $  B(E2)  $ for 2$^{ + }_{1}$ ($\Delta$=2) to 0$^{ + }_{2}$ 
($\Delta$=0)
is zero. Experimentally $  B(E2,2^{+}_{1} \rightarrow 0^{+}_{2})  $ = 48$^{ + 75}_{-20}$ e$^{2}$ 
fm$^{4}$
compared to $  B(E2,2^{+}_{1} \rightarrow 0^{+}_{1})  $ = 96(16) e$^{2}$ fm$^{4}$
(see Table I of \cite{el19}).
An improved half-life for the 0$^{ + }_{2}$ is important since it
helps to determine the $\Delta$ mixing.

One of the key experiments for $^{32}$Mg is the two-neutron
transfer from $^{30}$Mg$  (t,p)  $, where the first two 0$^{ + }$ states
were observed with about equal strength \cite{wi10}.
This observatiom has proven difficult to understand
(see the references in \cite{ma16}).
Starting from a $\Delta$=0 configuration for the
$^{30}$Mg ground state, one
can populate the $\Delta$=0, 0$^{ + }$ configuration in $^{32}$Mg by
$  (sd)^{2}  $ transfer and the $\Delta$=2, 0$^{ + }$ configuration
by $  (fp)^{2}  $ transfer.
Macchiavelli et al., \cite{ma16}  analyzed the $  (p,t)  $
cross sections by used centroid energies for
the $\Delta$=0,2,4 configurations of 1.4, 0.2 and 0.0 MeV,
respectively, obtained with the SDPF-U-MIX
Hamiltonian \cite{sdpfumix}. This three-level model could account for
the experimental observation with a ground state
that is 4\% $\Delta$=0, 46\% $\Delta$=2 and 40\% $\Delta$=4
together with a ground state wavefunction for $^{30}$Mg that
has 97\% $\Delta$=0 and 3\% $\Delta$=2.
In this three-level model for $^{32}$Mg, the main part of the
$\Delta$=0 configuration is in the 0$^{ + }_{3}$ state predicted
to be near 2.2 MeV (See Table I in \cite{ma16}).

Two-proton knockout from
$^{34}$Si provides more information. Starting with
a pure $\Delta$=0 configuration for the $^{34}$Si ground-state,
only $\Delta$=0, 0$^{ + }$ configurations in $^{32}$Mg can be made.
In the two-proton knockout experiment
of  \cite{ki21}, \cite{ki22}, strong 0$^{ + }$ strength is observed
in the sum of the first two 0$^{ + }$ states (Fig. 9 of \cite{ki22}). The strength
to the 0$^{ + }_{1}$ and 0$^{ + }_{2}$ states cannot be separated due to
the long lifetime of the 0$^{ + }_{2}$ state.
Significant  strength to 0$^{ + }$ states above 1.5 MeV was not observed,
in contradiction to that predicted in the
three-level model above \cite{ma16} or the SDPF-M model.
More needs to be done to understand the structure of $^{32}$Mg
and how it connects to the experimental data discussed above.

\begin{figure}
\includegraphics[scale=0.5]{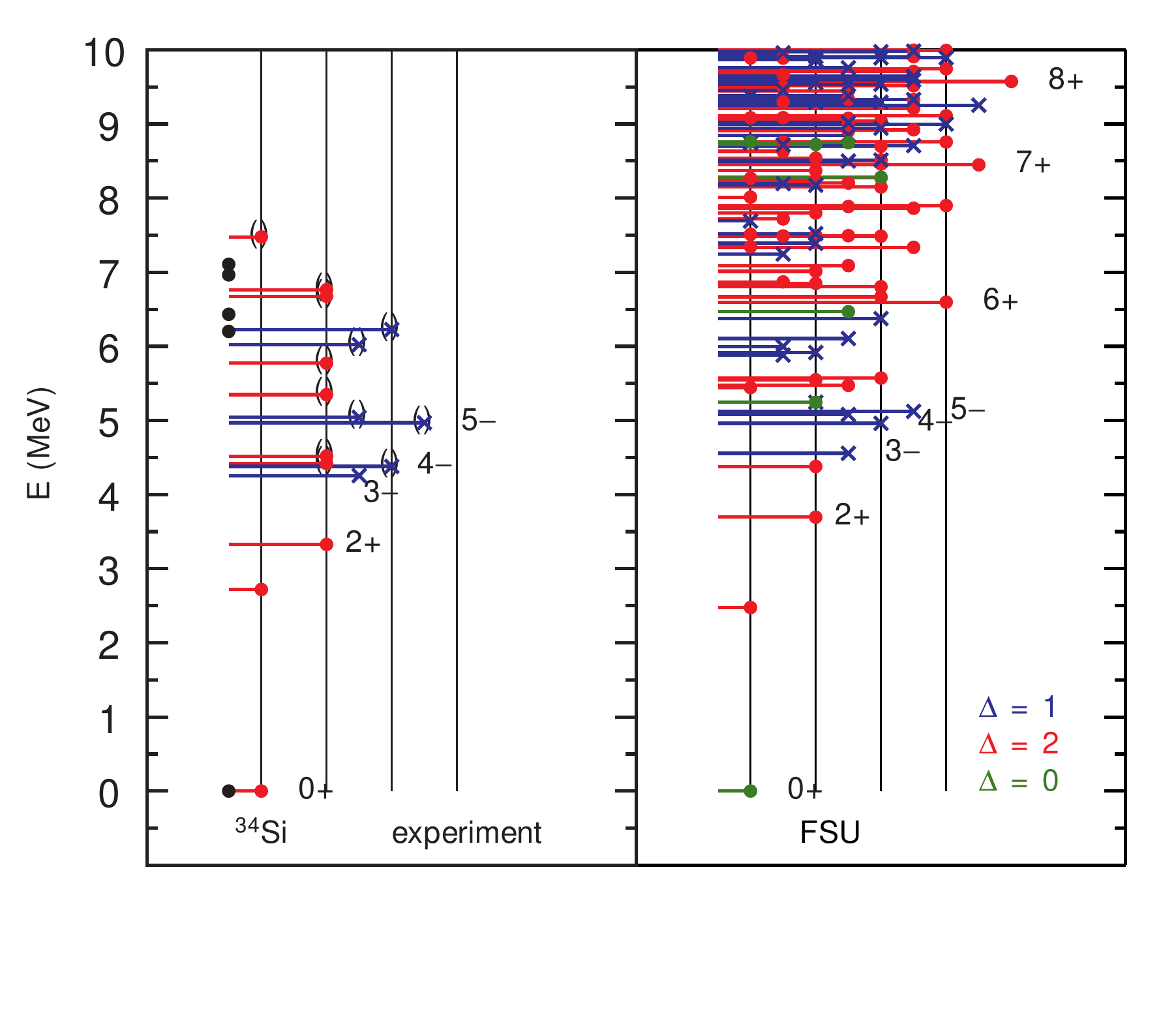}
\caption{Spectrum of $^{34}$Si obtained with the FSU Hamiltonian \cite{fsu}
compared to experiment.
The length of the horizontal lines are proportional to the $  J  $.
The experimental parity is indicated by blue for negative parity
and red for positive parity. Experimental $  J^{\pi }  $ values
that are tentative are shown by (), and those with multiple
of no $  J^{\pi }  $ assignments are shown by the black points.
The calculated results are obtained with the FSU Hamiltonian with pure $  \Delta   $ configurations.
The parities are positive  for $  \Delta =0  $ (green) and $  \Delta =2  $ (red) and
negative for $  \Delta =1  $ (blue).
}
\label{ (5) }
\end{figure}

\begin{figure}
\includegraphics[scale=0.6]{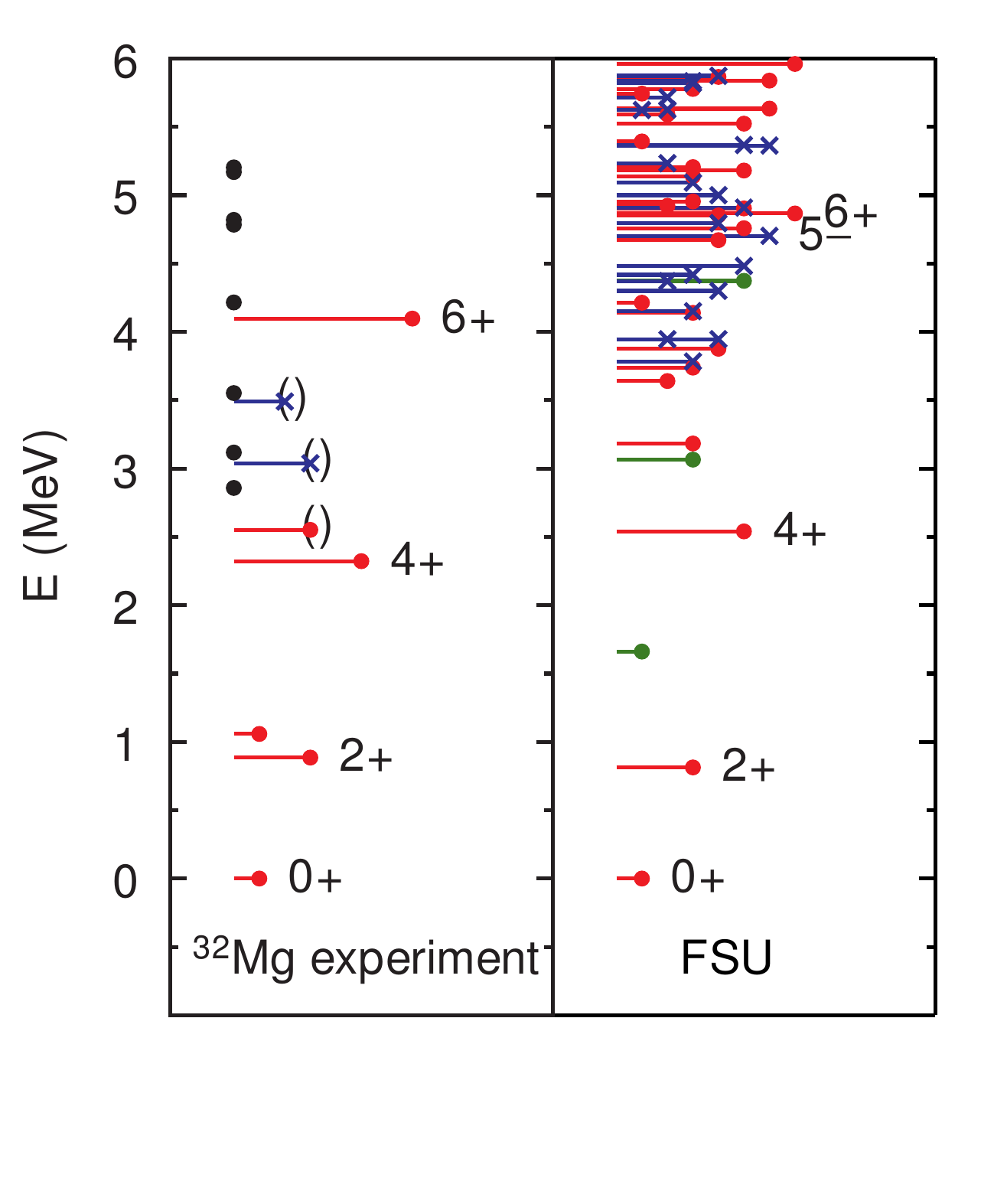}
\caption{Spectrum of $^{32}$Mg obtained with the FSU Hamiltonian \cite{fsu}.
The results are obtained with pure $  \Delta   $ configurations.
The spins are proportional to the length of the horizontal lines.
The parities are positive  for $  \Delta =0  $ (green) and $  \Delta =2  $ (red) and
negative for $  \Delta =1  $ (blue).
}
\label{ (6) }
\end{figure}

\begin{figure}
\includegraphics[scale=0.6]{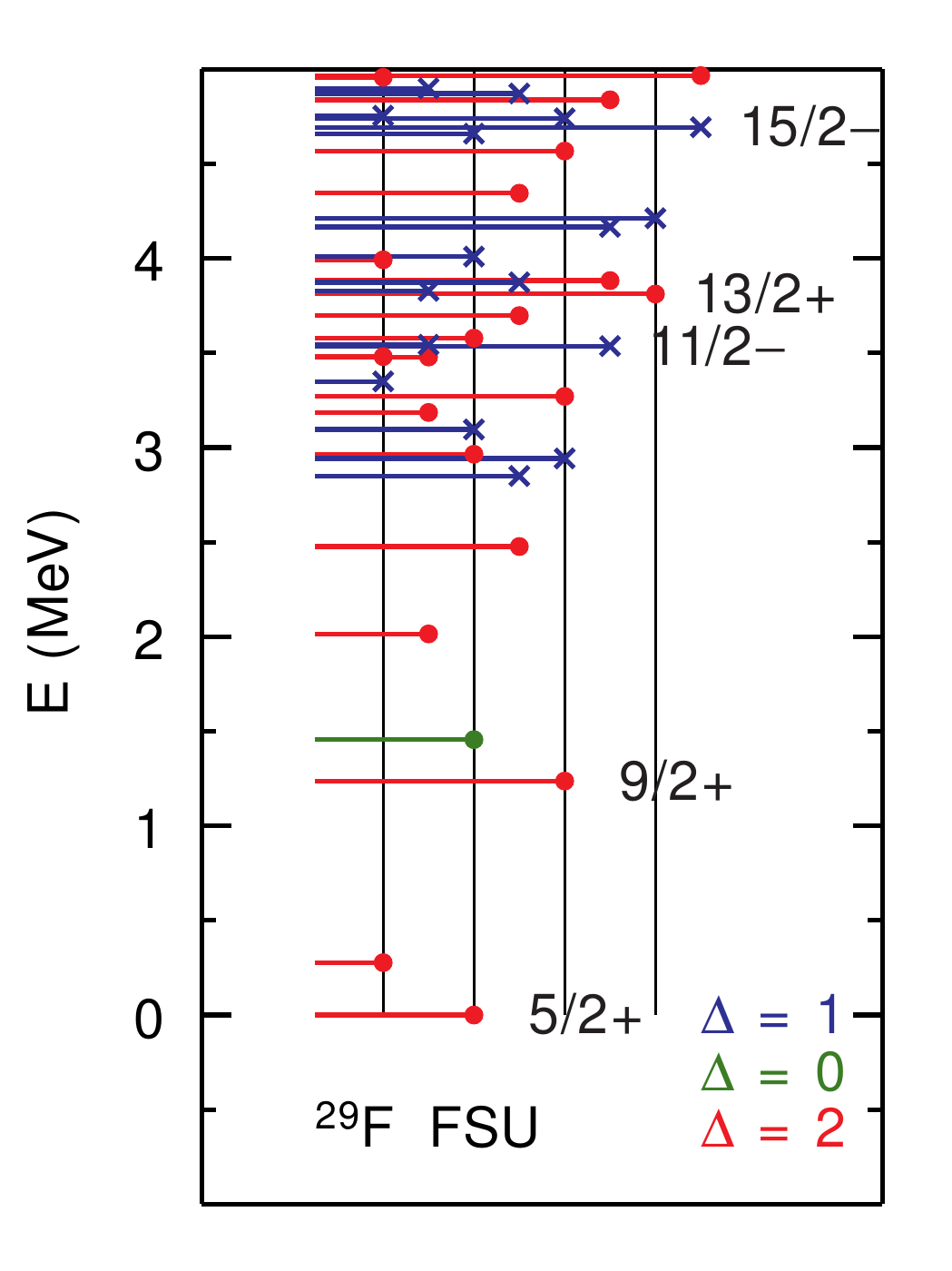}
\caption{Spectrum of $^{29}$F obtained with the FSU Hamiltonian \cite{fsu}.
The results are obtained with pure $  \Delta   $ configurations.
The spins are proportional to the length of the horizontal lines.
The parities are positive  for $  \Delta =0  $ (green) and $  \Delta =2  $ (red) and
negative for $  \Delta =1  $ (blue).
}
\label{ (7) }
\end{figure}

Results from the FSU Hamiltonian
provide an extrapolation down to $^{28}$O.
$^{29}$F has been called a "lighthouse on the island-of-inversion" \cite{fo20}.
The FSU results for $^{29}$F are shown in Fig. (7).
The lowest state is 5/2$^{ + }$ with  a $  \Delta =2  $ configuration.
The lowest 1/2$^{ + }$, 3/2$^{ + }$, 7/2$^{ + }$ and 9/2$^{ + }$  $\Delta$=2 states
are dominated by the configuration with $  0d_{5/2}  $ coupled
to the $\Delta$=2, 2$^{ + }$ state in $^{28}$O at 1.26 MeV. (The
$  0d_{5/2}  $ coupled to 2$^{ + }$, 5/2$^{ + }$ configuration
is spread over many higher 5/2$^{ + }$ states in $^{29}$F.)
The $  \Delta =3  $ states for $^{29}$F start at 3.9 MeV.
An excited state in $^{29}$F at 1.080(18) MeV \cite{do17}
made from proton knockout from $^{30}$Ne
was suggested to be 1/2$^{ + }$ on the basis comparisons to the SDPF-M
calculations shown in \cite{do17}.

With the FSU Hamiltonian, for $^{27}$F the lowest $\Delta$=0, 5/2$^{ + }$ state is  1.9 MeV below
the $\Delta$=2, 5/2$^{ + }$ state.
The large FSU occupancy of 1.38 in $^{29}$F for the loosely bound
$  0p_{3/2}  $ orbital may explain the observed neutron halo \cite{ba20}.
In particular, TNA[(0p$_{3/2}$)]$^{=}$0.62 for the $^{29}$F, $\Delta$=2, 5/2$^{ + }$ ground
state going to the $^{27}$F, $\Delta$=0, 5/2$^{ + }$ ground state.
Improved mass mesurements are needed
for the neutron-rich fluorine and neon isotopes.

Results for these calculations depend on the
ESPE extrapolation down to $^{28}$O contained in the
FSU interaction.
The ESPE for the neutron orbitals as a function of $  Z  $
obtained with the FSU Hamiltonian with ($\Delta$=0) are shown in Fig. (8)
(for $^{34}$Si I assume a $  (0d_{5/2})^{6}  $ configuration for the protons).
These are compared with the results from the Skx EDF calculations \cite{skx}.

For unbound states, the energies can be approximated by first
increasing the EDF central potential to obtain a wavefunction bound by,
for example, 0.2 MeV, and
then taking the expectation value of the wavefunction value with original EDF Hamiltonian.
This method provides a practical
approximation to the centroid energy. Results for the unbound resonances
could be calculated more exactly from neutron scattering on the EDF potential.

The results in Fig. (8) show that the $  N=20  $ shell gap decreases
from about 7.0 MeV in $^{34}$Si to about 2.7 MeV in $^{28}$O.
The major part of this decrease is due to the lowering
energy for $  1p_{3/2}  $ relative to $  0f_{7/2}  $ as the states
become more unbound. The energies for these two states
cross around $  Z=10  $.
Recent experimental information on the ESPE near $^{28}$Mg and their interpretation
similar to those of Fig. (8)
with a Woods-Saxon potential is given in \cite{mac21}.
For the FSU Hamiltonian,
the loose-binding effects are implicitly built into the
monopole components of the TBME from the SVD fit to
data on the BE and excitations energies.

There is also an increase in the gap
in $^{34}$Si due to the tensor force contribution to the
spin-orbit splitting that is built into the FSU Hamiltonian.
The spin-orbit tensor force is zero in the double-$  LS  $ closed
shell nuclei $^{28}$O and $^{40}$Ca.

The $  fp  $ ESPE obtained from the Skx EDF  \cite{skx}
from $^{30}$Ne to $^{78}$Ni are shown in Fig. (9).
The energies of $  1p  $ and $  0f  $ systematically shift due to the
finite-well potential.

\begin{figure}
\includegraphics[scale=0.6]{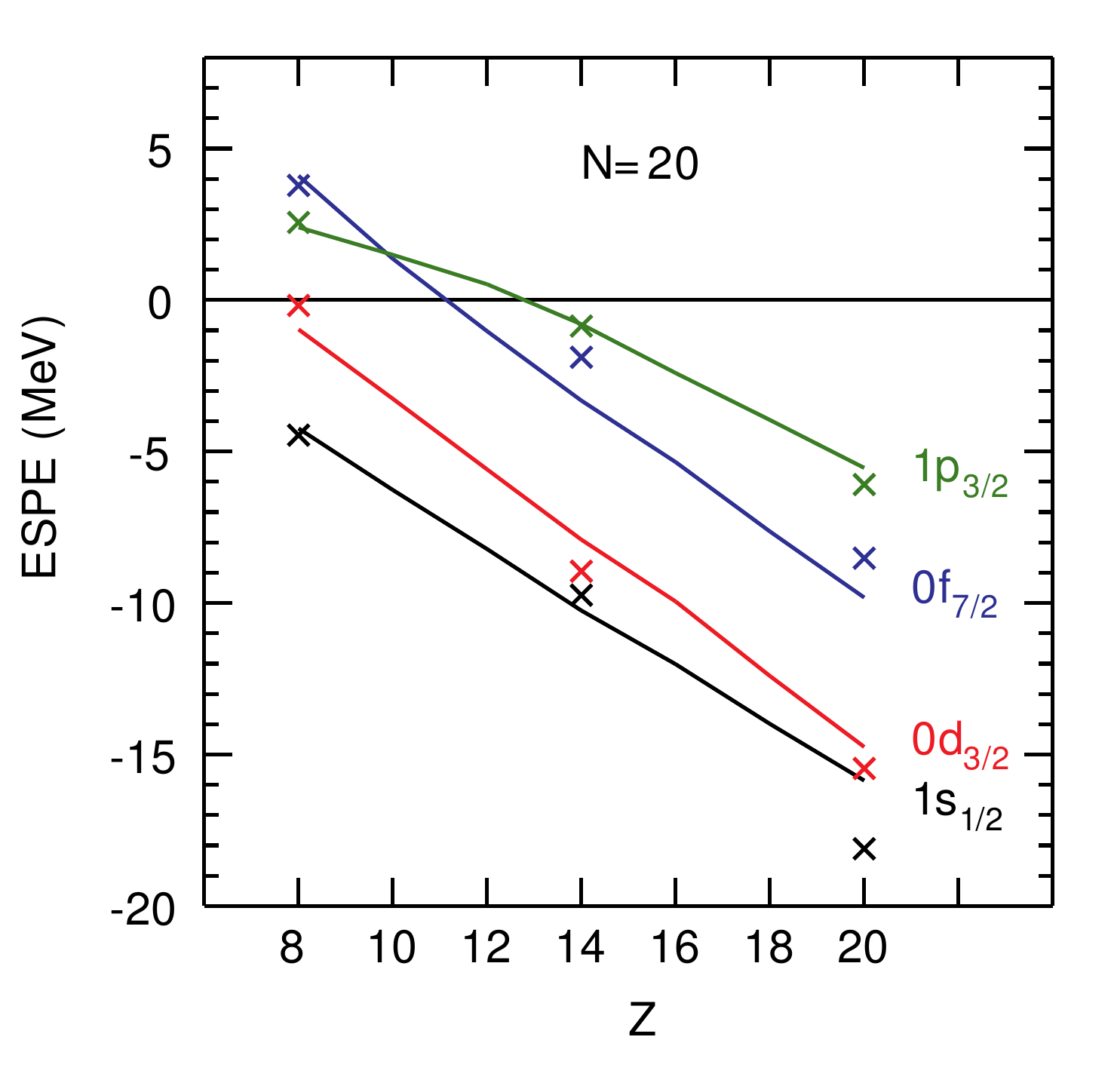}
\caption{ESPE for neutron orbitals
as a function of proton number.
The lines are from the Skx EDF \cite{skx} calculations.
The crosses are from the FSU \cite{fsu} Hamiltonian
calculations.
\label{ (8) }
}
\end{figure}

\begin{figure}
\includegraphics[scale=0.6]{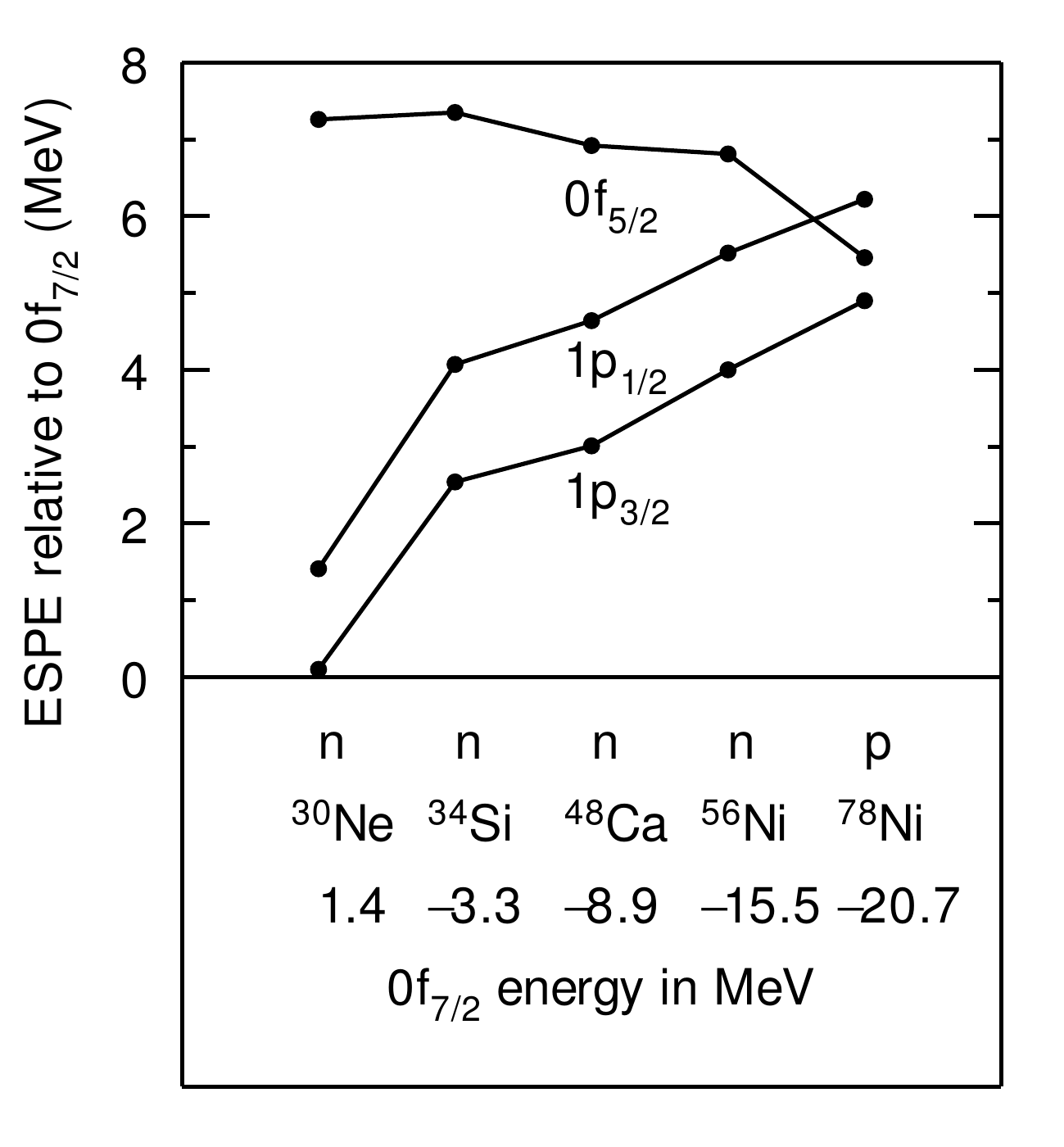}
\caption{ESPE for the $  fp  $ proton (p)
and neutron (n) orbitals obtained from the
Skx EDF interaction \cite{skx} for a range of nuclei.
\label{ (9) }
}
\end{figure}

For nuclei near the neutron drip line, there are
few bound states that can be studied by their gamma decay.
States above the neutron separation energy neutron decay. These
neutron decays can be complex both experimentally
and theoretically. The neutron decay spectrum depends upon how
the unbound states are populated. They are often
made by proton and neutron knockout reactions.
For one- and two-nucleon knockout one can calculate spectroscopic
factors that can be combined with a reaction model to find
which states are most strongly populated. A recent example of
this type of calculation was for two-proton knockout from $^{33}$Mg going to
$^{31}$Ne \cite{ch21}. One neutron decay can often go to excited states
in the daughter \cite{ch21}. And multi-neutron decay can occur. It is
important to measure the neutrons in coincidence with the final nucleus
and its gamma decays. On the theoretical side one must use
the calculated wavefunctions to obtain neutron decay spectra.

An example of multi-neutron decay is in the
one-proton knockout from $^{25}$F to make $^{24}$O \cite{th03}, \cite{ta20}.
The calculated one-proton knockout spectroscopic factors
showed that $  0d_{5/2}  $
knockout mainly leads to the ground state of $^{24}$O, and
that $  0p  $
knockout leads to many negative-parity states
above the neutron separation energy of $^{24}$O. These excited states
multi-neutron decay to $^{21-23}$O \cite{th03}. However, in the
(p,2p) reaction \cite{ta20} it was suggested from the
momentum-distribution of $^{23}$O that a low-lying
positive-parity excited state in $^{24}$O above the neutron separation energy
was strongly populated by $  0d  $ removal,
in strong disagreement with the calculations of \cite{th03}.
This experimental result should be confirmed.

The two-neutron decay of $^{26}$O has a remarkably small $  Q  $ value of 0.018(5) MeV \cite{o25}.
The theoretical $  Q  $ value from USDC Hamiltonian \cite{usdc} is 0.02(15) MeV.
The decay width depends strongly on the $\ell$ for the $  \ell ^{2}  $ two-nucleon decay amplitude.
From Fig. 2b of \cite{gr11} pure $  \ell ^{2}  $ two-nucleon decays widths
with the experimental $  Q  $ value are
approximately, 10$^{-4}$, 10$^{-8}$ and 10$^{-14}$ MeV, for $\ell$=0, 1 and 2, respectively.
The calculated two-neutron transfer amplitudes (TNA) in the $  sd  $ model space with the
USDC Hamiltonian
are 0.99 for $  (0d_{3/2})^{2}  $ and 0.16 for $  (1s_{1/2})^{2}  $.
Thus $  \Gamma  = [{\rm TNA}(1s_{1/2})^{2}]^{2} \, \Gamma _{sp}(Q) \approx 0.003  $ keV.
The $  (1p_{3/2})^{2}  $ TNA will be on the order of $  <(0d_{3/2})^{2}\mid V\mid 
(1p_{3/2})^{2}>/2\Delta E  $,
where $  \Delta E  $ is the energy difference between the the $  1p_{3/2}  $ and
$  0d_{d3/2}  $ states in $^{25}$O.
With typical values of $  <(0d_{3/2})^{2}\mid V\mid (1p_{3/2})^{2}> \approx 2  $ MeV and
$  \Delta E \approx 2  $ MeV \cite{le15}
giving TNA=0.5, the $  (1p_{3/2})^{2}  $ contribution to the two-neutron decay width
will be small.

The nucleus $^{28}$O is unbound to four neutron decay.
The theoretical understanding of this complex decay involves the four-body continuum
\cite{gr11}. These continuum calculations strongly depend upon the single-particle states
involved (see Fig 2d in \cite{gr11}).
With the FSU Hamiltonian,
the $  \Delta =2  $ configuration for $^{28}$O lies 0.8 MeV below the $  \Delta =0  $ (closed-shell)
configuration due to the pairing correlations.
The calculated four-neutron decay energy is 1.5 MeV.
The energy should be lowered by an explicit treatment of the many-body continuum.
Thus,
the "island-of-inversion" may be  a "peninsula of inversion" extending
from $^{32}$Mg all the way to the neutron drip line.
(Later I discuss what may be the first true "island-of-inversion" between $^{60}$Ca
and $^{78}$Ni.)  There are many paths for the four-neutron decay of $^{28}$O.
For example, in the FSU $  \Delta =2  $  model, it may proceed by a
relatively fast $  (1p_{3/2})^{2}  $ decay to the $^{26}$O ground state followed by its
decay to $^{24}$O.

\section{The region of $^{42}$Si}

We will compare results
for two widely used effective Hamiltonians for this model space, SDPF-MU \cite{sdpfmu}
and SDFP-U-SI \cite{sdpfu}, together with those based on the IMSRG method \cite{st21}.
The MU and USI Hamiltonians are "universal" in the sense that a single
Hamiltonian with a smooth mass-dependence is applied to a wide mass region.
MU is used for all nuclei in this model space, while USI was designed for
$  Z \leq 14  $ (the SDPF-U version was designed for $  Z>14  $ \cite{sdpfu}).

The 2$^{ + }$ energy in $^{42}$Si $  [Z,N]=[14,28]  $ (0.74 MeV) is low compared to
those in $^{34}$Si [14,20] (3.33 MeV) and $^{48}$Ca [20,28] (3.83 MeV).
$^{34}$Si and $^{48}$Ca are doubly-magic due to the $  LS  $ magic number 20.
$^{28}$Si [14,14] has a well-known intrinsic oblate deformation \cite{mo21}.

The 2$^{ + }$ energy in $^{20}$C [6,14] (1.62 MeV) is low compared to
those in $^{14}$C [6,8] (7.01 MeV) and $^{22}$O [8,14] (3.20 MeV).
$^{14}$C and $^{22}$O are doubly-magic due to the
$  LS  $ magic number 8.
Hartree-Fock calculations \cite{sa04} as well as CI
calculations for the $  Q  $ moment within the $  p-sd  $ model space \cite{pe11}
show that $^{12}$C and $^{20}$C have intrinsic oblate shapes.

The oblate shapes for $^{28}$Si and $^{42}$Si are shown by
their $  E2  $ maps in Fig. (10) and Fig. (11).
The transition from spherical to oblate shapes for the $  jj  $ doubly-magic
numbers can be qualitatively understood in the Nilsson diagram
as shown, for example, for $^{42}$Si in Fig. (12).
The highest filled Nilsson orbitals have rather flat
energies between $\beta$=0 and $\beta$= -0.3.
The important aspect is the concave bend of the
2$\Omega^{ \pi }$ [N,n$_{z}$,$\Lambda$] = 1$^{ + }$ [2,2,0] proton
and 1$^{-}$ [3,3,0] neutron Nilsson orbitals for oblate shapes.
For the heavier $  jj  $ doubly-magic nuclei,
$\ell$ increases and the $  j=\ell +1/2  $ orbital
decreases in energy, the bend will not be so large and the
energy minima come closer to $\beta$=0.
This is illustrated in Fig. (10). In panels (b) and (c)
the $  0d  $ spin-orbit gap is small enough to give an oblate rotational
pattern. The oblate shape is manifest in the
positive $  Q  $ moments. In panel (a) the $  0d  $ spin-orbit
gap is increased by one MeV and the rotational energy  pattern is
broken. The pattern in panel (a) is similar to that obtained
for $^{56}$Ni in the $  fp  $ model space as shown in Fig. (13).
An interesting feature for $^{56}$Ni is the relatively strong
 0$^{ + }_{2}$ to 2$^{ + }_{1}$ $  B(E2)  $. I am not aware of a
simple explanation for this.

The oblate bands in $^{28}$Si and $^{42}$Si
are linked to the $  0d_{5/2}  $ and $  0f_{7/2}  $ orbitals.
For completeness, I show the $  E2  $ maps for
$^{12}$C and $^{20}$C obtained with the WBP Hamiltonian \cite{wbp}
in Fig. (14). For these nuclei the oblate ground-state
bands are linked with
the $  0p_{3/2}  $ and $  0d_{5/2}  $ orbitals.

\begin{figure}
\includegraphics[scale=0.6]{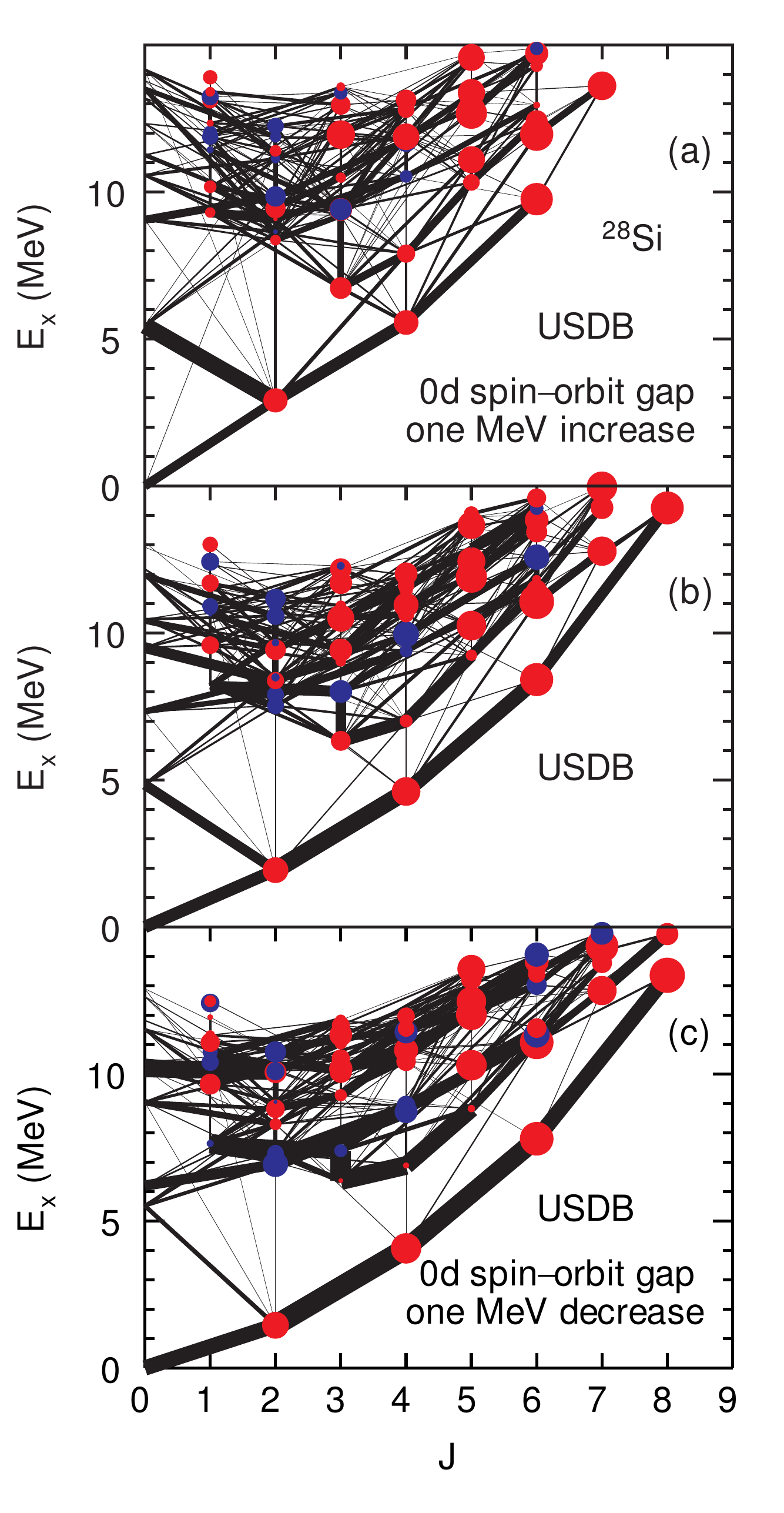}
\caption{$  E2  $ maps for $^{28}$Si.
For each $  J  $ value ten states were calculated.
The widths of the lines are proportional to the B(E2).
Lines for B(E2) less than 5\% of the largest value are not shown.
The radius of the circles are proportional to Q$_{sp}$.
To set the scale for this and the
following figures, the 2$^{ + }_{1}$ to 0$^{ + }_{1}$ B(E2) = 82 e$^{2}$ fm$^{4}$ = 1.51 WU,
and Q(2$^{ + }_{1}$) = $+$19 e fm.
The results in panel (b)
were obtained with
the USDB Hamiltonian in the $  sd  $ model space.
The results in panel (a) were obtained with the
$  0d  $ spin-orbit energy gap increased by one MeV,
and the results in panel (c) were obtained with the
$  0d  $ spin-orbit energy gap decreased by one MeV.
}
\label{ (10) }
\end{figure}

\begin{figure}
\includegraphics[scale=0.6]{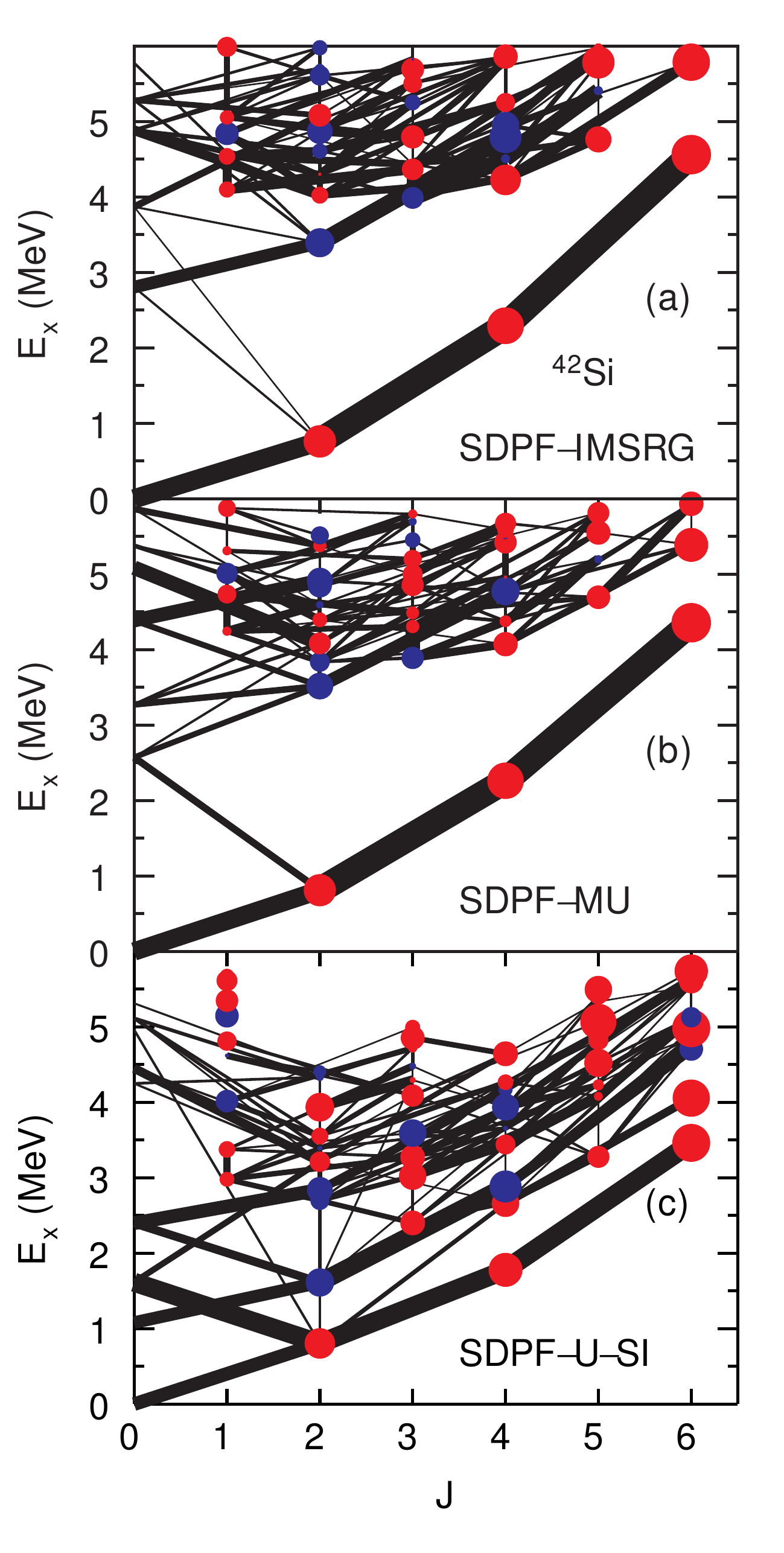}
\caption{$  E2  $ maps for $^{42}$Si obtained with
the three Hamiltonians SDPF-IMSRG \cite{st21}, SDPF-MU \cite{sdpfmu}, and SDPF-I-SI
\cite{sdpfu}.
}
\label{ (11) }
\end{figure}

\begin{figure}
\includegraphics[scale=0.5]{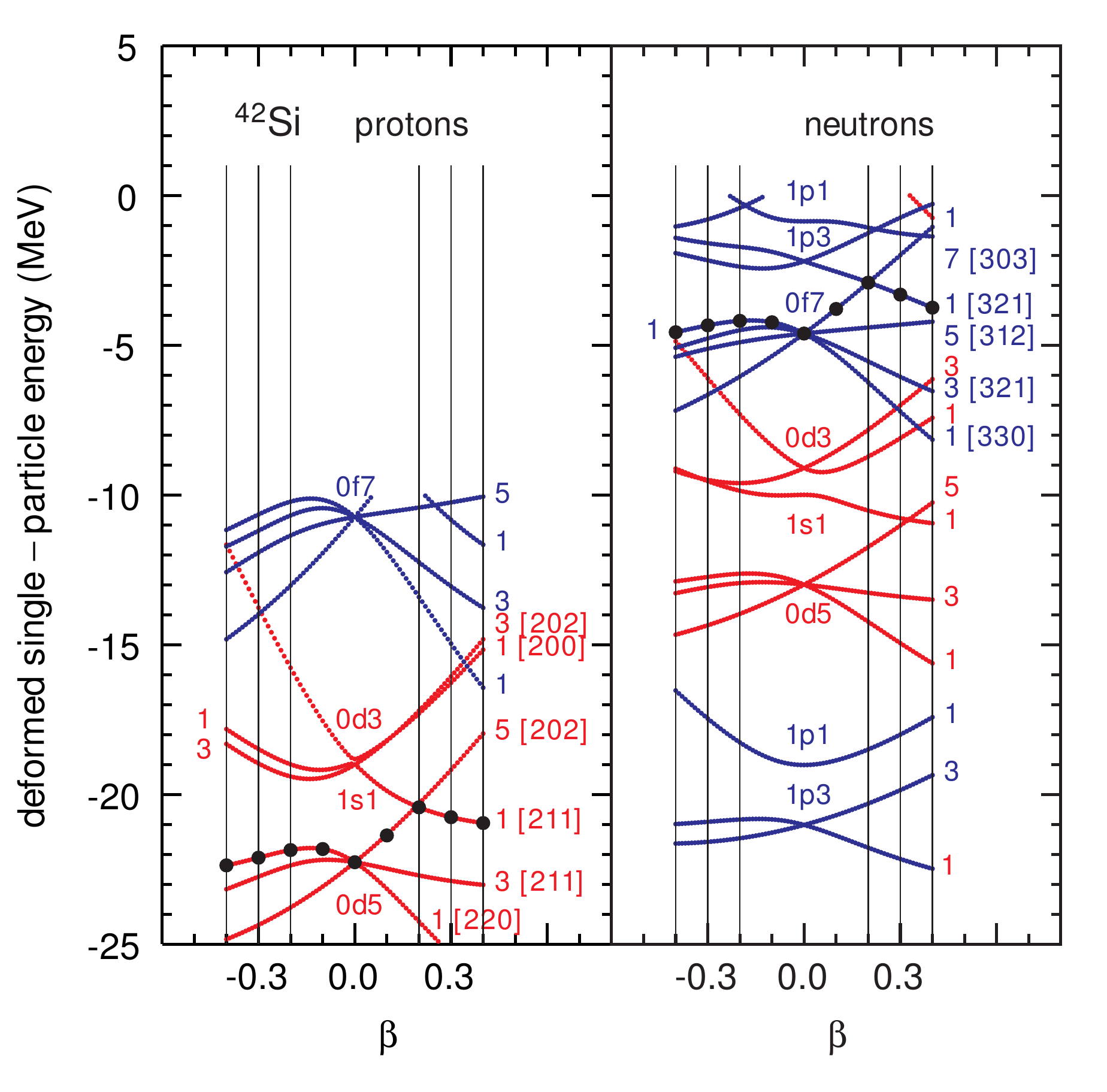}
\caption{Nilsson diagram for $^{42}$Si.
At $\beta$=0 the orbitals are labeled by $  (n,\ell ,2j)  $, and at larger
deformation they are labeled the Nilsson quantum numbers 2$\Omega$ [N,n$_{z}$,$\Lambda$].
The parity is shown by the blue (negative) and red (positive) lines.
The black dots show the
highest Nilsson states occupied as a function of deformation $\beta$.
This is made using the computer code WSBETA \cite{dws} with the
potential choice ICHOIC=3. I reduced the spin-orbit potential
for protons to make the spherical energies for the
$  0d_{3/2}  $ and $  1s_{1/2}  $ orbitals at about the same.
}
\label{ (12) }
\end{figure}

\begin{figure}
\includegraphics[scale=0.6]{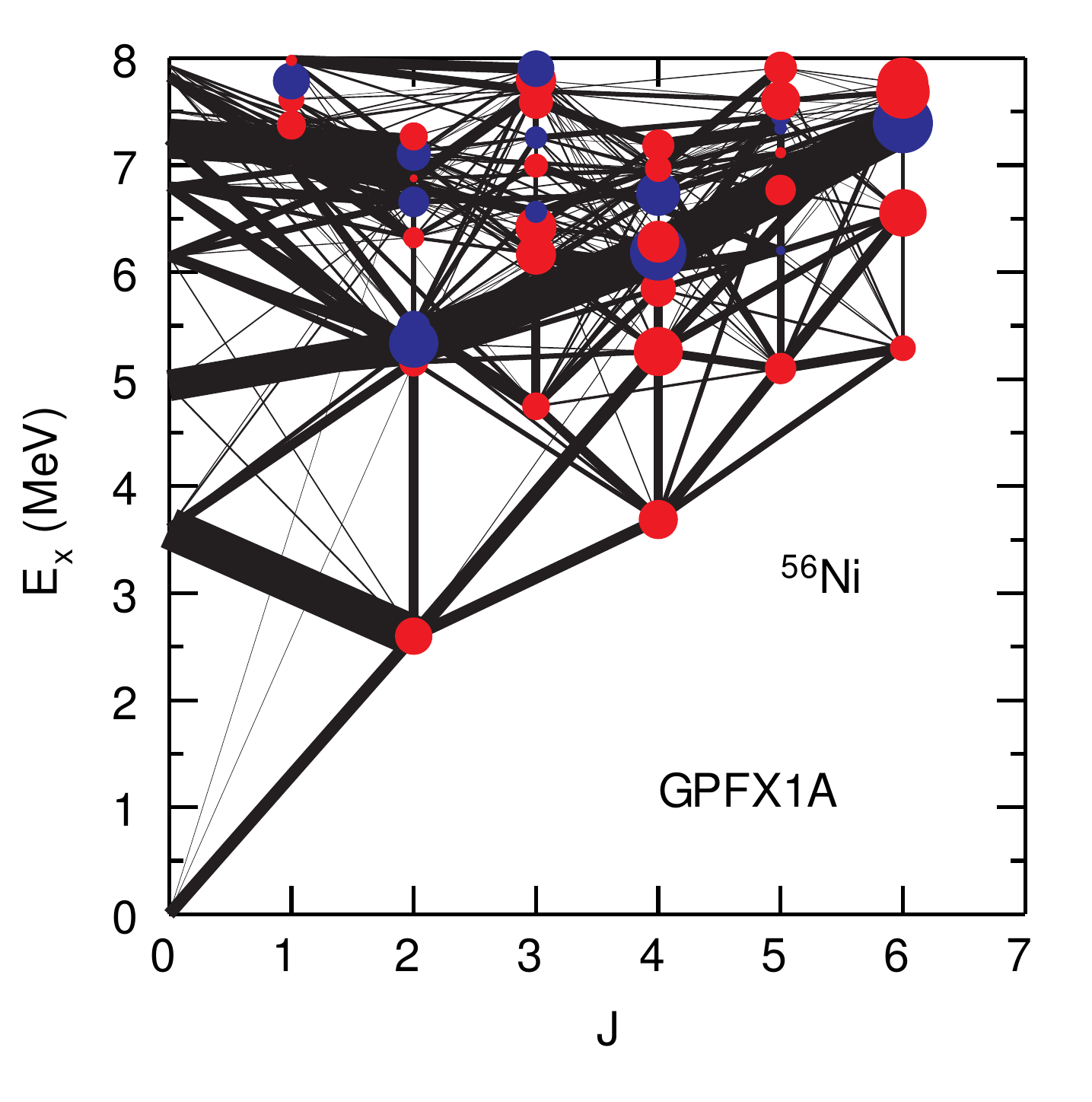}
\caption{$  E2  $ map for $^{56}$Ni obtained with
the GPFX1A Hamiltonian \cite{ho04}, \cite{enam04} in the full $  fp  $ model space.
}
\label{ (13) }
\end{figure}

\begin{figure}
\includegraphics[scale=0.6]{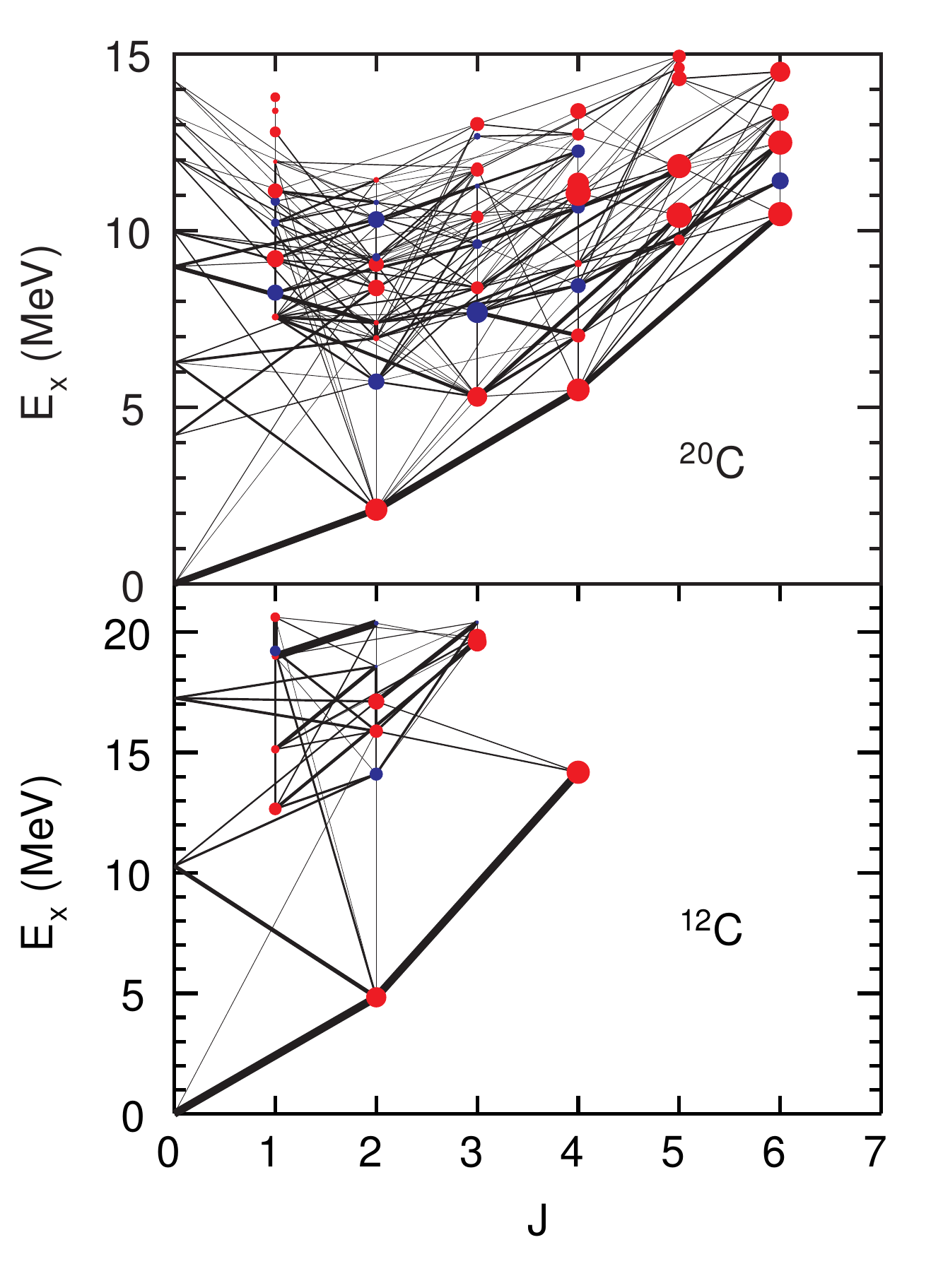}
\caption{$  E2  $ maps for $^{12}$C and $^{18}$C
obtained with the WBP Hamiltonian \cite{wbp}.
}
\label{ (14) }
\end{figure}

For CI calculations the B(E2) depend on the effective charge parameters
$  e_{p}  $ and $  e_{n}  $.  In the harmonic-oscillator basis the
$  E2  $ operator connects states within a major shell as well
as those that change $  N_{o}  $ by two. The $  E2  $ strength
function contains low-lying $  \Delta N_{o}=0  $ strength as well
"giant-quadrupole" strength near an energy of 2$\hbar\omega$.
The effective charges account for the renormalization of the
proton and neutron components of the
$  E2  $ matrix elements within the CI basis of a major shell due to admixtures of
the $  1p-1h  $, $  \Delta N_{o}=2  $  proton configurations.
For the calculations I show here,
I use effective charges which depend on the
model space. They are chosen to best reproduce
observed B(E2) values and quadrupole moments within that model space.
These are the $  sd  $ model space with $  e_{p}=0.45  $ and $  e_{n} = 0.36  $ \cite{ri08},
the $  fp  $ model space with $  e_{p}=e_{n}=0.50  $ \cite{ho04}
and the neutron-rich $  sd-pf  $ model space with $  e_{p}=e_{n}=0.35  $ \cite{sdpfmu}.
Since low-lying excitations in nuclei are mostly isoscalar,
only $  e_{p}+e_{n}  $ is well determined. It takes
special situations such as a comparison of B(E2) in mirror
nuclei \cite{ri04} to obtain the isovector combination $  e_{p}-e_{n}  $.

The isoscalar effective charge decreases for more neutron-rich nuclei
(e.g. the drop from 0.5 in the $  fp  $ model space to
0.35 in the $  sd  $ model space).
This can be understood by the macroscopic model of Mottelson
(pages 507-555 of \cite{bm2}),
by the microscopic Hartree-Fock calculations of Sagawa et al. \cite{sa04},
and by the microscopic models discussed in \cite{br77}, \cite{lo21}.
Microscopic models also give an orbital dependence to the
effective charge. A recent example of this is for
the relatively small B(E2) value for the
the 1/2$^{ + }$ to 5/2$^{ + }$ transition in $^{21}$O \cite{he20}. This
transition
is dominated by the $  1s_{1/2}-0d_{5/2}  $ $  E2  $ matrix
element,
and the relatively small neutron effective charge
is due to the node in the $  1s_{1/2}  $ wavefunction.

The results for CI calculations for $^{42}$Si are shown in Fig. (11) for
three Hamiltonians.
The IMSRG Hamiltonian is based on a VS-IMSRG calculation \cite{st21}
similar to that used for \cite{ma21}.
The interpretation of the spectroscopic quadrupole moments, $  Q_{s}  $,
shown in Fig. (11) in terms of an
intrinsic shape $  Q_{o}  $ is given by the rotational formula \cite{lo70}
$$
Q_{s} = \frac{2K^{2} - J(J+1)}{(J+1)(2J+3)} Q_{o} \, {\rm e},       \eqno({2})
$$
with $  K=0  $ for the ground state bands in even-even nuclei.
The MU \cite{sdpfmu} and IMSRG \cite{st21} calculations
show an intrinsic oblate ground-state band, ($  Q_{s}>0  $ and $  Q_{o}<0  $), followed by
a large energy gap to other more complex states.
The U-SI Hamiltonian \cite{sdpfu} also gives an oblate ground-state
band, but there is also an intrinsic prolate band at relatively low energy.
The presence of this low-lying prolate band dramatically increases the
level density below 4 MeV \cite{to13}, \cite{si42}.

The Nilsson diagram in  Fig. (12) shows a higher energy prolate
minimum related to a crossing of the
1$^{-}$ [3,2,1] and 7$^{-}$ [3,0,3] Nilsson orbitals near $\beta$ = $+$0.3.
At present
there is not enough experimental information to
determine the energy of the prolate band in $^{42}$Si.
The structure of $^{42}$Si is a touchstone
for understanding all of the nuclei near the
drip line in this mass region.
More complete experimental results for the energy
levels of $^{42}$Si are needed.
The low-lying structure of $^{42}$Si depends on the
details of the neutron ESPE that are affected by the continuum
for the $  0p  $ orbitals. The deformed neutron ESPE need to
be established by one-neutron transfer reactions on $^{42}$Si.

Deformation for $  N=28  $ as a function of $  Z  $ is determined by
how the proton Nilsson orbitals are filled in Fig. (12). When
six protons are added to make
to make $^{48}$Ca with $  Z=20  $, there is a sharp energy minimum for protons
at $\beta$=0, and thus $^{48}$Ca is doubly-magic. For $^{44}$S
the protons have a intrinsic prolate minimum near $\beta$=$+$0.2 where
the neutrons are near the crossing of the 2$\Omega^{ \pi }$=1$^{-}$ and 7$^{-}$ orbitals 
\cite{ut15}.
In $^{44}$S a K=4$^{ + }$
isomer at 2.27 MeV coming from the two quasi-particle state
made from these two neutron orbitals was observed \cite{s44}.
In $^{43}$S rotational bands associated with these two $\Omega$ states
have been observed \cite{s43}. All of these features
are reproduced by CI calculations based on the SDPF-MU \cite{sdpfmu}
and SDPF-U \cite{sdpfu} Hamiltonians. At higher excitation energy,
the CI energy spectra are more complex than anything that
could be easily understood by the collective model.

\begin{figure}
\includegraphics[scale=0.6]{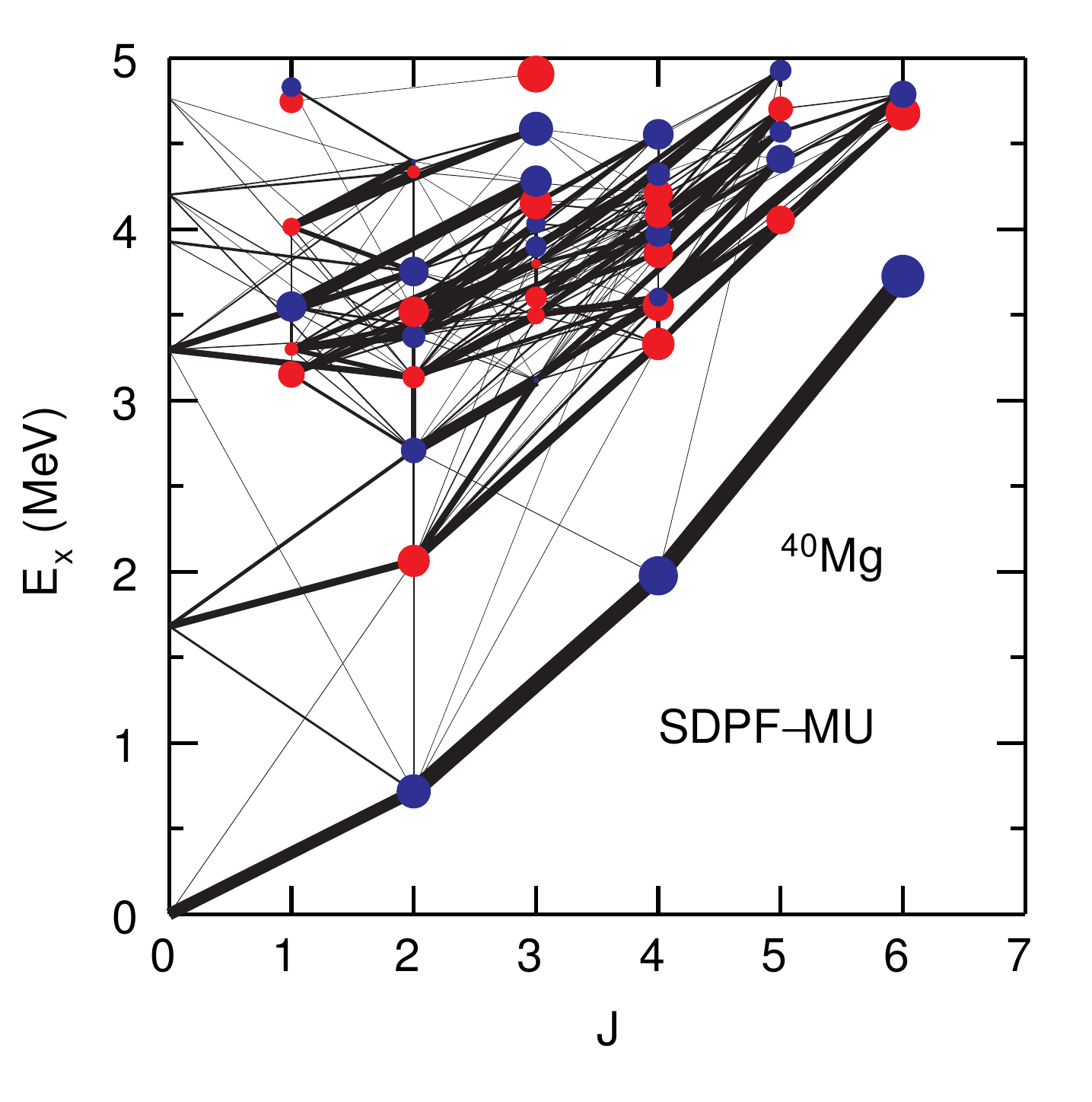}
\caption{$  E2  $ map for $^{40}$Mg obtained with
the SDFPF-MU Hamiltonian.
}
\label{ (15) }
\end{figure}

The $  E2  $ map obtained with the SDPF-MU Hamiltonian for $^{40}$Mg
is shown in Fig. (15).
In this case the ground state band has an intrinsic prolate shape.
In the nuclear chart prolate shapes are most common \cite{sh18},
in contrast to the oblate shapes obtained for  $  jj  $ magic numbers
discussed above.  The oblate shape for $^{40}$Mg can be understood in the Nilsson
diagram of (12). When two protons are removed, the energy
minimum for protons shifts to positive $\beta$ in the 3$^{-}$ [2,1,1] orbital.
The experimental energy of the first 2$^{ + }$ is 500(14) keV \cite{cr19}
compared to the result of 718 keV obtained with the SDPF-MU Hamiltonian.
Models that explicitly include the $  \ell =1  $ levels in the continuum are needed.

\section{The region of $^{60}$Ca}

Many Hamiltonians have been developed for the calcium isotopes
for the $  fp  $ model space. Near $^{42}$Ca it is well known
that $\Delta$=2 $  sd-pf  $  proton excitations are necessary for the low-lying
intruder states and their mixing with the $  fp  $ configurations
which greatly increase the B(E2) values compared to those
obtained in the $  fp  $ model space \cite{lo21}.
In the doubly-magic nucleus
$^{48}$Ca, the $  sd-pf  $ intruder states start with the
0$^{ + }$ state at 4.28 MeV  \cite{br98}. The $^{48-55}$Ca nuclei
exhibit low-lying spectra which are dominated by $  fp  $ configurations \cite{ma21}.
There are weak magic numbers at $  N=32  $ and 34 as shown in Fig. (2).
The reason for the low value of the pairing for the $  1p_{1/2}  $
at $  N=33  $ was discussed in \cite{br98}.

The KB3G \cite{kb3g} and GPFX1A \cite{ho04}, \cite{enam04} Hamiltonians have provided
predictions for the spectra in this region which have been
a source of comparison for many experiments over the last 20 years.
Both of these are "universal" Hamiltonians for the $  pf  $ model space.
Recently it has been shown that a data-driven Hamiltonian for the
calcium isotopes improves the description of all of the known data
\cite{ma21}.
This is called the UFP-CA (universal $  fp  $ for calcium) Hamiltonian.
All of the known energy data for $  N \geq 28  $ can be
described by an SVD-derived  Hamiltonian that is close to the
starting IMSRG Hamiltonian for $^{48}$Ca.
UFP-CA is able to describe the energy data for $  N \geq 28  $
with an rms error of 120 keV. In particular, the calculated
$  D(N)  $ values shown by the red line in Fig. (2) agree
extremely well
with the data (the black points).

The UFP-CA Hamiltonian does not explicitly involve the $  2s-1d-0g  $
orbitals, but the influence of these orbitals are present
in their contributions to the renormalization into the $  fp  $ model space.
This renormalization is contained microscopically in the
IMSRG starting point, as well as empirically in the SVD fit.

The success of UFA-CA is similar to the success of the USD-type
Hamiltonians in the $  sd  $ model space for all nuclei except those
in the island-of-inversion. If the UFP-CA predictions for $^{55-59}$Ca
turn out to be in relatively good agreement with experiment, the
implication is that $^{60}$Ca will be a doubly-magic nucleus
similar to that of $^{68}$Ni \cite{ma21}. If that is the case,
$^{60}$Ca will be
the last doubly-magic nucleus to be discovered.
In \cite{ma21} EDF models were used to estimate the $  0f_{5/2}  $
$  0g_{9/2}  $
shell gap at $  N=40  $ to be about about 3.0 MeV. The
implication of this for $  D(N)  $ is shown by the red
dashed line in Fig. (2).
The $  0g_{9/2}  $ orbital will first appear as intruder states
in the low-lying spectra of $^{55-60}$Ca.
These nuclei can be reached by proton knockout on the
scandium and titatium isotopes. The proton knockout will
be dominated by $  0f_{7/2}  $ removal to the low-lying
$  fp  $ neutron configurations. An example of this
is the population of the ground state of $^{54}$Ca
from $^{55}$Sc \cite{br21}. Protons will also be removed from the
$  1s_{1/2}  $ and $  0d_{3/2}  $ orbitals to populate
states at higher energy such as the negative parity
state in $^{54}$Ca. These will mix with the $  2s-1d-0g  $ configurations
and neutron decay to the lighter calcium isotopes.
For example, in $^{57,59}$Ca a 9/2$^{ + }$ ($  0g_{9/2}  $) state
just above the $  S_{n}  $ value would neutron decay to the
(0$^{ + }$, 2$^{ + }$, 4$^{ + }$) multiplet predicted in $^{56,58}$Ca (see Fig. 1 in \cite{ma21}).
Calculations that include proton excited from $  sd  $ to $  pf  $
and neutrons excited from $  pf  $ to $  sdg  $ will be
needed to understand the neutron dacays of these states.

The position of the $  0g_{9/2}  $ orbital is crucial
for the structure of nuclei around $^{60}$Ca \cite{bh20}.
Lenzi et al. \cite{le10} have extrapolated the neutron
effective single-particle energies from $  Z=28  $ down to $  Z=20  $ based on their LNPS
Hamiltonian. Their $  0f_{5/2}  $-$  0g_{9/2}  $ ESPE gap for $^{60}$Ca is close to zero
(see Fig. 1 in Ref. \cite{le10}) and the structure of $^{60}$Ca is
dominated by $\Delta$=4 ($  fp  $ to $  sdg  $) configurations (see table I in \cite{le10}).
With LNPS, $^{60}$Ca is very different from $^{68}$Ni which is dominated by the closed $  fp  
$-shell
configuration ($\Delta$=0).
Below $^{68}$Ni, the nuclei $^{66-70}$Fe,\cite{sa15} $^{64-66}$Cr \cite{sa15} and $^{62}$Ti 
\cite{cortes20},
 have deformed
spectra coming from $  fp-sdf  $ island-of-inversion for $  N=40  $.
Calculations with the  LNPS  Hamiltonian \cite{le10} show that these are all dominated by
$\Delta$=4.
The $  N=40  $ island-of-inversion is the topic of another
contribution to this series of papers \cite{ga21}.

The existence of $^{60}$Ca, confirmed only recently,
agrees with UFP-CA as well as with most of the other predictions \cite{ta18}.
It will be exciting to have more complete experimental data for nuclei around $^{60}$Ca from
FRIB and other radioactive-beam facilities.

\section{The region of $^{78}$Ni}

$^{78}$Ni has recently been established as a double-$  jj  $ magic nucleus
from the relatively high energy of 2.6 MeV for the  2$^{ + }_{1}$ state \cite{ni78}.
More detailed magic properties can be obtained from the
$  D(N)  $
and $  D(Z)  $ derived from new experiments on the
masses around $^{78}$Ni. The ESPE can be established
from the masses together with the low-lying spectra
of $^{77}$Ni, $^{79}$Ni, $^{77}$Co and $^{79}$Cu.
A proton knockout experiment from $^{80}$Zn  has recently been
used to establish excitation energies of low-lying states
in $^{79}$Cu \cite{ni79}. In particular, the ground state
and two lowest-lying states are likely associated with
the triplet of states shown in Fig. (9). In comparison with the
extrapolations of CI calculations shown in \cite{ni79},
the order is likely to be $  0f_{5/2}  $, $  1p_{3/2}  $ and $  1p_{1/2}  $.
The single-particle nature of low-lying states around $^{78}$Ni will
require one-nucleon transfer experiments.

The position of the proton $  0g_{9/2}  $ orbital above $^{78}$Ni
is important for Gamow-Teller strength in the electron-capture
rates for core-collapse supernovae similations \cite{za19}, \cite{ti19}.
The filling of the $  0g_{9/2}  $ orbital
leads to $^{100}$Sn on the proton drip line.
$^{100}$Sn has the largest calculated B(GT)
value (see Table A1 in \cite{st21}) due to nearly filled $  0g_{9/2}  $
orbital decaying into the nearly empty $  0g_{7/2}  $ orbital.
The understanding of $^{100}$Sn \cite{lu19} and other nuclei
near the proton drip line in this mass region will
be improved by radioactive-beam experiments.

As shown in Fig. 4b of \cite{ni78}, large scale CI calculations
predict a deformed band with $  \beta  \approx +0.3  $ around 2.6 MeV.
$^{56}$Ni is also spherical with a 2$^{ + }_{1}$ state observed at 2.7 MeV.
For $^{56}$Ni the deformed band is predicted to start at 5.0 MeV
as shown in (13). The relatively low-lying deformed
band in $^{78}$Ni is predicted to lead to a "5th island-of-inversion"
in $^{76}$Fe and other nuclei with $  N=50  $ below $  Z=28  $ \cite{no16}.

\section{Conclusions}

I have discussed the new physics related to the properties
of nuclei near the drip lines that will be studied
by the next generation of rare-isotope beam experiments.
In, particular I have focused on four "outposts"
for the regions of $^{28}$O, $^{42}$Si, $^{60}$Ca and $^{78}$Ni
where new experiments will have the greatest impact
on our understanding.

\section{Acknowledgements}

I acknowledge support from National Science Foundation grant
PHY-2110365. I thank Ragnar Stroberg for providing the VS-IMSRG
Hamiltonian for $^{42}$Si.

\end{document}